\documentclass[11pt,a4]{article}

\usepackage[dvipdfmx]{graphicx}
\usepackage{amssymb}
\usepackage{unit}
\usepackage{color}
\usepackage{mathrsfs}
\usepackage{amsmath}
\usepackage{subcaption}
\usepackage{color}
\usepackage{amsthm}
\usepackage{amscd}
\usepackage{algorithm}
\usepackage{algorithm,algpseudocode}

\theoremstyle{definition}

\begin{document}
\title{%
Synthesis of Spatial Charging/Discharging Patterns\\
of In-Vehicle Batteries for Provision of Ancillary Service\\
and Mitigation of Voltage Impact
}%
\author{
Naoto Mizuta, Yoshihiko Susuki, Yutaka Ota, and Atsushi Ishigame\footnote{N. Mizuta, Y. Susuki, and A. Ishigame are with Department of Electrical and Information Systems, Osaka Prefecture University, Japan (\texttt{susuki@eis.osakafu-u.ac.jp}). 
Y. Ota is with Department of Electrical and Electronics Engineering, Tokyo City University, Japan.
}
}%
\date{November 2018}

\maketitle

\begin{abstract}
We develop an algorithm for synthesizing a spatial pattern of charging/discharging operations of in-vehicle batteries for provision of Ancillary Service (AS) in power distribution grids. 
The algorithm is based on the ODE (Ordinary Differential Equation) model of distribution voltage that has been recently introduced. 
In this paper, firstly, we derive analytical solutions of the ODE model for a single straight-line feeder through a partial linearization, thereby providing a physical insight to the impact of spatial EV charging/discharging to the distribution voltage. 
Second, based on the analytical solutions, we propose an algorithm for determining the values of charging/discharging power (active and reactive) by in-vehicle batteries in the single feeder grid, so that the power demanded as AS (e.g. a regulation signal to distribution system operator for primary frequency control reserve) is provided by EVs, and the deviation of distribution voltage from a nominal value is reduced in the grid. 
Effectiveness of the algorithm is established with numerical simulations on the single feeder grid and on a realistic feeder grid with multiple bifurcations. 
\end{abstract}
 
\section{Introduction}

Large-scale integration of Electric Vehicles (EVs) is now expected to contribute to the future power grid operation \cite{Ipakchi}. 
In a transmission grid, a large population of in-vehicle batteries is being investigated for the so-called Demand Response (DR) that aims to shift the peak load and to provide regulation supports for primary and secondary (load) frequency controls \cite{Sousa,Wu}. 
In a distribution grid, the so-called \emph{Distribution System Operator} (DSO) \cite{Arriaga} as a load dispatching center or aggregator is investigated for managing EVs in order to conduct the DR. 
The sort of these system-level ideas is generally termed as \emph{Ancillary Service} (AS) \cite{Rebours} provided by EVs and has been studied by many groups of researchers: see e.g. \cite{Kempton,Tomic,Ota}. 

The present paper addresses the provision of AS for primary frequency control reserve, called frequency response in PJM \cite{Rebours}.  
It is required to achieve fast responsiveness of several seconds to several minutes \cite{Kempton}. 
{\color{black}If the fast response is lacking, then some sorts of frequency reserves have to be considered as secondary reserves and not a primary one.}   
In-vehicle batteries are capable of responding faster than synchronous generators used in large thermal power plants and are hence suitable for the frequency response.  
{\color{black}The authors of \cite{Marinelli_JES7} report that in their experiment with three EVs in campus microgrid, the overall response delay of EVs following a change of frequency is in the ranges of two to three seconds, which is sufficient as the requirement of primary reserve listed in \cite{Rebours}. 
Regarding} 
this, in this paper we will propose a computationally simple method for determining the values of charging/discharging power by in-vehicle batteries in order to satisfy the power demanded as AS.  

There are several problems on the provision of AS by EVs. 
One problem is to manage the impact of charging/discharging of a large number of EVs to a distribution grid. 
An EV is regarded as an autonomously moving battery in the spatial domain and can conduct the charging and discharging (in principle) \emph{anywhere} in the grid. 
Thus, to maintain the nominal distribution voltage while providing the AS by EVs is a challenging subject. 
In fact, Clement-Nyns \emph{et al.} \cite{Clement1,Clement2} study the impact of EV charging to the distribution grid using the load-flow analysis and propose optimization-based methods for determining the timing and amount of charging power in order to reduce the voltage deviation. 
Also, Falahi \emph{et al.} \cite{Falahi} study the control of reactive power by grid-connected inverters of in-vehicle batteries for the voltage regulation and propose an optimization-based method for determining the reactive power injected to a grid from EVs in order to reduce the voltage deviation. 

Currently, we are developing methodology and tools for provision of AS under cooperation with an EV-sharing system, as a part of the research project ``Integrated Design of Local EMSs\footnote{Energy Management Systems} and their Aggregation Scenario Considering Energy Consumption Behaviors and Cooperative Use of Decentralized In-Vehicle Batteries \cite{Itoh}." 
An \emph{EV-sharing system} has a function of tracking trajectories of EVs to allocate them upon user's request \cite{Fairley,Tan}. 
The authors of \cite{Kawashima:CCTA17} propose to use the function of EV-sharing system for community-level energy management. 
This work opens a new possibility of integrated transportation-energy management, in which the EV-sharing system operator and DSO co-work for provision of AS.  
This integration is a typical System-of-Systems (SoS) application, which is addressed in the focus of this journal, and in this paper we address a design problem on the physical layer of energy transmission in the SoS.  
Here, in the physical layer of transportation, it is planned that a large population of shared EVs is distributed in a relatively-small area, referred to as the \emph{Last Mile Transportation} within a range of a few kilometers \cite{Tan}. 
Thus, a large number of charging stations can be densely placed in the small area (see \cite{Autolib} as an ongoing project in Paris, France) and possibly causes a non-negligible voltage impact.  

The purpose of this paper is to develop an algorithm for synthesizing a spatial pattern of charge/discharge operations of in-vehicle batteries for not only provision of AS but also mitigation of voltage impact. 
This algorithm is based on the ODE (Ordinary Differential Equation) representation of distribution voltage derived recently by Chertkov \emph{et al.} \cite{Chertkov}. 
This representation keeps spatial information of balanced distribution grids and is hence direct to the synthesis of spatial pattern of charging/discharging operations. 
The ODE representation is exploited in \cite{Susuki2} for assessing the impact of spatio-temporal EV charging to a distribution grid, in which practically-measured data on movements of shared EVs are used. 
For the above purpose, we intend a situation where a DSO works as a provider of primary frequency control reserve by coordinated use of EVs, receives a regulation signal from an upper transmission system operator, and then utilizes the algorithm for provision of AS and management of distribution voltage. 

The contributions of this paper are two-fold. 
One is to derive analytical solutions for distribution voltage for a single, straight-line feeder model. 
This derivation becomes possible with the ODE representation, and the derived solutions explicitly describe the position-dependent profile of voltage and thus provide a clear physical insight to the impact of spatial EV charging/discharging. 
The other contribution is based on the physical insight and to establish an algorithm for determining values of charging/discharging power by in-vehicle batteries in a manner such that power demanded as AS is provided by EVs, and the deviation of distribution voltage from a nominal value is reduced. 
This algorithm is easy to implementation, needs no iterative computation like optimization with standard power-flow equations, and is expected to be fast and scalable in terms of the grid's complexity. 
This advantage is suitable to the current motivation on providing the primary frequency control reserve from shared EVs.   
The algorithm proposed here is demonstrated with numerical simulations of the two models: the single feeder model and a realistic feeder model with multiple bifurcations. 

This paper is a substantially-enhanced version of our conference paper \cite{Mizuta}. 
The algorithm in \cite{Mizuta} was developed for synthesizing the active power only; in this paper we generalize it for synthesizing both active and reactive power. 
According to this, we provide new simulation results for both the single feeder model and the realistic feeder model. 
The latter model does not appear in \cite{Mizuta}. 
The rest of the paper is organized as follows. 
Section~\ref{sec:ODE} introduces the ODE representation of distribution voltage profile. 
In Section~\ref{sec:analysis} we derive analytical solutions for the distribution voltage profile. 
In Section~\ref{sec:design} we propose the synthesis algorithm based on the analytical solutions. 
The algorithm is numerically evaluated in Section~\ref{sec:demo}. 
Section~\ref{sec:conclusion} concludes this paper with a brief summery and future directions.

\section{ODE Representation of Distribution Voltage Profile}
\label{sec:ODE}

\begin{figure}[t]
\centering
\includegraphics[width=.65\hsize]{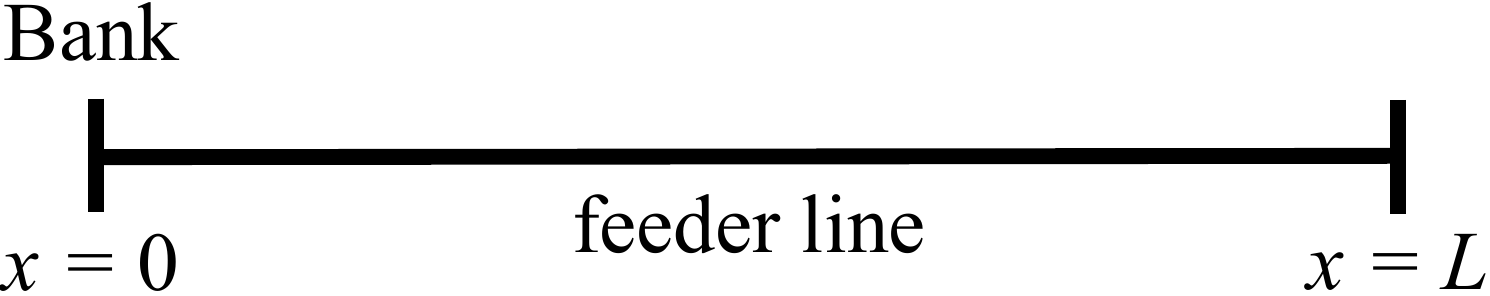}
\caption{%
Single, straight-line distribution feeder that starts at a bank, through a finite-length line with length $L$, and ends at a non-loading terminal. 
}%
\label{fig:feeder1}
\end{figure}

First of all, we introduce the ODE representation of distribution voltage profile based on \cite{Chertkov}. 
For simplicity of the introduction, we assume that no voltage regulation device such as load ratio control transformer and step voltage regulator is operated. 
Thus, we can consider the voltage profile starting at a distribution substation (bank) that is continuous in space (length). 
Note that it is possible to include the effect of such voltage regulation devices in the ODE representation: see \cite{Susuki2}. 
Now consider a single, straight-line, balanced distribution feeder shown in Figure~\ref{fig:feeder1}, starting at a bank where we introduce the origin of the one-dimensional displacement (location) $x\in\mathbb{R}$ as $x=0$. 
The voltage phasor at the location $x$ is represented with $v(x)\exp\{\ii\theta(x)\}$, where $\ii$ denotes the imaginary unit, $v(x)$ the \emph{voltage amplitude} [$\rm V$], and $\theta(x)$ the \emph{voltage phase} [$\rm rad$]. 
Also, as new functions in $x$, define the \emph{power transfer density} [$\rm V^2/km$] at $x$ by $s(x):=-v(x)^2d\theta(x)/dx$ and the \emph{voltage gradient} [$\rm V/km$] by $w(x):=dv(x)/dx$. 
Thus, the four functions $\theta, v, s$, and $w$ are related with the following first-order nonlinear ODE \cite{Chertkov}:
\begin{equation}
\left.
\begin{array}{ccl}
\displaystyle \frac{d\theta}{dx} &=& \displaystyle -\frac{s}{v^2}~~~(v\neq 0),
\\\noalign{\vskip 2mm}
\displaystyle \frac{dv}{dx} &=& w,
\\\noalign{\vskip 2mm}
\displaystyle \frac{ds}{dx} &=& \displaystyle \frac{b(x)p(x)-g(x)q(x)}{g(x)^2+b(x)^2},
\\\noalign{\vskip 1mm}
\displaystyle \frac{dw}{dx} &=& \displaystyle \frac{s^2}{v^3}-\frac{g(x)p(x)+b(x)q(x)}{v\{g(x)^2+b(x)^2\}}.
\end{array}
\right\}
\label{eqn:ode}
\end{equation}
The two functions $g(x)$ and $b(x)$ in (\ref{eqn:ode}) are the position-dependent conductance and susceptance per unit-length [$\rm S/km$] and assumed to be continuous in $x$. 
Also, the function $p(x)$ (or $q(x)$) is the active (or reactive) power flowing into the feeder (note that $p(x)>0$ indicates the positive active-power flowing to the feeder at $x$). 
In what follows, we will call $p(x)$ and $q(x)$ the \emph{power density functions} in [$\rm W/km$] and [$\rm Var/km$]. 

Here, to solve the nonlinear ODE (\ref{eqn:ode}), it is necessary to determine not only the functions $g(x)$, $b(x)$, $p(x)$, and $q(x)$ but also a \emph{boundary condition}. 
Consider again the single distribution feeder in Figure~\ref{fig:feeder1}. 
At the starting point $x=0$, due to voltage regulation at the bank, we naturally set $v(0)$ and $\theta(0)$ to be constant. 
Throughout this paper, $v(0)$ coincides with unity in per-unit system and $\theta(0)$ with zero as a reference. 
At the end point $x=L$ (namely, no feeder and load exist at $x>\color{black}L
$), by supposing that the end is not loaded, we have the conditions $s(L)=0$ and $w(0)=0$ \cite{Susuki2}. 
The above boundary condition is summarized as
\begin{equation}
\theta(0)=0,~~v(0)=1,~~s(L)=0,~~w(L)=0.
\label{eqn:BC}
\end{equation}

Note that the nonlinear ODE (\ref{eqn:ode}) is applicable to a distribution feeder with a bifurcation by incorporating an appropriate boundary condition \cite{Susuki2}, which will be used in Section~\ref{subsec:multi}. 
Static ZIP models \cite{Machowski} of loads can be also incorporated \cite{Susuki2}.\footnote{Dynamic models of loads such as induction machines can be connected to (\ref{eqn:ode}).  
In this case, the model becomes a system of partial differential equations, and thus its computation (numerical approximation of solutions of the model) should be carefully considered.}   
Furthermore, we numerically demonstrate in \cite{Susuki2} that the distribution voltage profiles computed with (\ref{eqn:ode}) and standard power-flow equations for a common feeder model are consistent. 
This implies that the choice of models for distribution voltage does not affect effectiveness evaluation of the proposed algorithm in Section~\ref{sec:demo}.

\section{Analytical Solutions for Distribution Voltage Profile}
\label{sec:analysis}

In this section, we derive analytical solutions of the voltage amplitude $v$ and gradient $w$ in the nonlinear ODE (\ref{eqn:ode}). 
These solutions provide a clear physical insight of the distribution voltage profile and lead to the synthesis of spatial charging/discharging patterns of in-vehicle batteries in the next section. 

We assume in this paper that both load and EV are represented as constant power model (PQ-bus type).  
Conventionally, grid-connected inverters for batteries are regulated at constant active power under unity power factor (implying zero reactive power)\cite{Yang}. 
Based on this, the proper power-flow model for batteries corresponds to PQ-bus type. 
Therefore, in the nonlinear ODE model, any in-vehicle battery is represented by constant power model (PQ-bus type) at the point of connection to a feeder. 

Now consider a case in Figure~\ref{fig:feeder1} where multiple charging stations as well as constant power loads are connected. 
Suppose that $N$ number of stations and loads are located at $x=\xi_i\in(0,L)$ ($i=1,\ldots,N$) satisfying $\xi_{i+1}<\xi_{i}$. 
Then, by denoting as $P_i$ (or $Q_i$) the active-power discharged / charged (consumed) (or the reactive-power supplied / consumed) at $x=\xi_i$, the power density functions $p(x)$ and $q(x)$ are given as
\begin{equation}
p(x)=\sum^{N}_{i=1}P_i\delta(x-\xi_i),~~~q(x)=\sum^{N}_{i=1}Q_i\delta(x-\xi_i) 
\label{eqn:pq}
\end{equation}
where $\delta(x-\xi_i)$ is the Dirac's delta function supported at $x=\xi_i$. 
For the current derivation, we assume that the conductance and susceptance of the feeder are constant in $x$: 
\[
g(x)=G,~~~b(x)=B 
\]
where $G$ and $B$ are constants. 
This is relevant when the single feeder is made of a common conductor. 
We also assume that all $v$ on the right-hand side of (\ref{eqn:ode}) are close to unity. 
Based on the two assumptions, the following approximation of the nonlinear ODE (\ref{eqn:ode}) is obtained: 
\begin{equation}
\left.
\begin{array}{ccl}
\displaystyle \frac{d\theta}{dx} &=& \displaystyle -s,
\\\noalign{\vskip 2mm}
\displaystyle \frac{dv}{dx} &=& w,
\\\noalign{\vskip 2mm}
\displaystyle \frac{ds}{dx} &=& \displaystyle \frac{B\cdot p(x)-G\cdot q(x)}{G^2+B^2},
\\\noalign{\vskip 2mm}
\displaystyle \frac{dw}{dx} &=& \displaystyle s^2-\frac{G\cdot p(x)+B\cdot q(x)}{G^2+B^2}.
\end{array}
\right\}
\label{eqn:approx}
\end{equation}
The differential equations are linear and can be solved for $p(x)$ and $q(x)$ given in (\ref{eqn:pq}). 
Then, the power transfer density $s(x)$ is derived from (\ref{eqn:BC}), (\ref{eqn:pq}), and (\ref{eqn:approx}) as the sum of charging (consumed) and discharging power along the portion of the feeder between $x$ and the end point as follows: 
\begin{eqnarray}
s(x)=
\left\{
\begin{array}{ll}
\displaystyle -\frac{1}{Z^2}\sum_{j\in\mathcal{I}_x}\left(BP_j - GQ_j\right), &x \neq \xi_i 
\\\noalign{\vskip 2mm}
\displaystyle -\frac{1}{Z^2}\sum_{j\in\mathcal{I}_{\xi_{i+1}}}\left(BP_j - GQ_j\right), &x = \xi_i - 0 
\\\noalign{\vskip 2mm}
\displaystyle -\frac{1}{Z^2}\sum_{j\in\mathcal{I}_{\xi_{i}}}\left(BP_j - GQ_j\right), &x = \xi_i + 0 
\end{array}
\right.
\label{eqn:s}
\end{eqnarray}
where $\mathcal{I}_x\subseteq\{1,2,\ldots,N\}$ is the set of all indexes $i$ satisfying $x<\xi_i$. 
Based on (\ref{eqn:BC}), (\ref{eqn:approx}), and (\ref{eqn:s}), the voltage gradient $w(x)$ is derived as follows: 
\begin{eqnarray}
\small
w(x)=
\left\{
\begin{array}{ll}
\displaystyle \frac{1}{Z^4}\sum_{j\in\mathcal{I}_x}\left(BP_{j} - GQ_j\right)^{2}(x-\xi_{i})
+\frac{1}{Z^2}\sum_{j\in\mathcal{I}_x}\left(GP_j + BQ_j\right) + f(\xi_i), &\xi_{i+1} < x < \xi_i 
\\\noalign{\vskip 2mm}
\displaystyle \frac{1}{Z^2}\sum_{j\in\mathcal{I}_{\xi_{i+1}}}\left(GP_j + BQ_j\right) + f(\xi_i), &x = \xi_i - 0 
\\\noalign{\vskip 2mm}
\displaystyle \frac{1}{Z^2}\sum_{j\in\mathcal{I}_{\xi_i}}\left(GP_j + BQ_j\right) + f(\xi_i), &x = \xi_i + 0 
\\\noalign{\vskip 2mm}
\displaystyle 0, &\xi_1 <x \leq L 
\end{array}
\right.
\label{eqn:w}
\end{eqnarray}
where $i\in\{1,\ldots,N\}$, $\xi_{N+1}:=0$, and
\begin{equation}
f(\xi_i):=\frac{1}{Z^4}\sum_{j\in\mathcal{I}_{\xi_i}}\sum_{k\in\mathcal{I}_{\xi_{j+1}}}\left(BP_{k} - GQ_k\right)^{\!2}
(\xi_{j+1}-\xi_{j}). 
\end{equation}
Consequently, the voltage amplitude $v(x)$ is derived from (\ref{eqn:approx}) and (\ref{eqn:w}) as in (\ref{eqn:v}). 
Here, because of $v(\xi_{N+1})=v(0)=1$, the term $v(\xi_{i+1})$ on the right-hand side in (\ref{eqn:v}) can be determined in a recursive manner. 
\begin{eqnarray}
v(x)=
\left\{
\begin{array}{ll}
\multicolumn{2}{l}{\displaystyle \frac{1}{2Z^4}\sum_{j\in\mathcal{I}_x}\left(BP_{j} - GQ_j\right)^{\!2}\left\{(x-\xi_{i})^{2}-(\xi_{i+1}-\xi_{i})^2\right\}}
\\\noalign{\vskip 2mm}
\displaystyle  +\frac{1}{Z^2}\sum_{j\in\mathcal{I}_x}\left(GP_j + BQ_j\right)(x-\xi_{i+1})
+f(\xi_i)(x-\xi_{i+1}) + v(\xi_{i+1}), &\xi_{i+1} \leq x\leq \xi_i,
\\\noalign{\vskip 2mm}
\displaystyle v(\xi_1), &\xi_1 < x \leq L. 
\end{array}
\right.
\label{eqn:v}
\end{eqnarray}

The derived analytical solutions provide physical implications for not only understanding the impact of EV charging/discharging but also synthesizing its spatial pattern for provision of AS. 
The voltage gradient $w(x)$ is a piecewise affine function in $x$, implying that the charging and discharging power by EVs affects the distribution grid linearly. 
Also, the solutions of $w(x)$ and $v(x)$ include the term of \emph{linear} sum of charging (consumed) and discharging power, $\sum_{j\in\mathcal{I}_x}(GP_j+BQ_j)$, uniquely determined by the location $x$. 
This clearly suggests that the voltage gradient $w(x)$ and amplitude $v(x)$ can be shaped with the charging and discharging power by EVs. 
Based on these observations, in the next section we will consider how to determine the values of charging and discharging power of in-vehicle batteries. 

The linear ODE (\ref{eqn:approx}) was derived for the single feeder with constant power model.  
The ``constant" implies that the amount of active and reactive power consumed is not dependent on voltage, current, and time.  
The analytical solutions (\ref{eqn:s}), (\ref{eqn:w}), and (\ref{eqn:v}) are thus not valid for situations in which a load is represented by static ZIP model or any dynamic model.  
Such situations need to be considered for evaluating the performance of AS frequency response as a controlled dynamic system \cite{Ota,Shimizu}, which is our future work.

\section{Synthesis of Charging/Discharging Pattern}
\label{sec:design}

In this section, based on the analytical solutions and physical insight in Section~\ref{sec:analysis}, we develop an algorithm for synthesizing the spatial pattern of charging/discharging power of in-vehicle batteries. 

\subsection{Synthesis Principle}
\label{sec:principle}

Our synthesis principle is based on the physical insight derived with the analytical solutions. 
By definition, the zero voltage gradient $w(x)=0$ implies no change of the voltage amplitude $v(x)$. 
We see from (\ref{eqn:w}) that $w(x)$ is piecewise {\color{black}affine} 
with respect to the position $x$, where there are multiple discontinuous points $x=\xi_i$ due to charging (consuming) or discharging operations of EVs and loads. 
Here, we focus on the terms including $\sum_{j\in\mathcal{I}_x}(GP_j + BQ_j)$ in (\ref{eqn:w}) for the synthesis. 
It should be stated at first that the magnitude of the position ($x$) dependent term in (\ref{eqn:w}) is much smaller than those of the other terms in normal setting of distribution grids \cite{Mizuta}. 
We thus see from (\ref{eqn:w}) that if $\sum_{j\in\mathcal{I}_x}(GP_j + BQ_j)$ is zero at $x \in (0,L)$, then $w(x)$ is approximately zero, that is, $v(x)$ does not change near $x$. 
Since the voltage amplitude $v(\xi_{N+1}=0)$ at the starting point is regulated at unity (nominal value) as shown in (\ref{eqn:v}) at $i=N$, shaping the voltage gradient $w(x)$ in a recursive manner from the end point of the feeder can reduce the deviation of voltage amplitude $v(x)$ from the nominal value. 
The reason of why we focus on $w(x)$ is that by comparison with $v(x)$, $w(x)$ clearly shows the grid's state, for example, loading condition, generating condition such as PV, and network topology, which will be shown in Figures~\ref{fig:simulation_4} and \ref{fig:rea_v_4}.\footnote{Also, regarding control, this focus on $w$ has a reasonable analogy with stabilization of the motion of a mass point in a potential field. 
It is known in control engineering that in order to stabilize the motion asymptotically with a proportional feedback law of force, the stabilization is never achieved with the displacement of point. 
However, the stabilization is possible with the velocity---the time derivative of displacement.}

Namely, the synthesis procedure is that we determine the active and reactive power supplied by EVs connected to each charging station such that $\sum_{j\in\mathcal{I}_x}(GP_j + BQ_j)$ approaches a value around zero. 
In a practical charging station, the amount of charging/discharging power is limited by State-Of-Charge (SOC) of in-vehicle batteries and inverter capacity of chargers. 
Also, for utilization of EVs connected to a distribution feeder, the loading capacity of bank transformer can limit the amount of provision of AS. 
In this paper, as introduced below, the loading capacity is sufficiently larger than the amount of active power demanded as AS, and hence we do not consider the loading capacity.   
When the calculated value of charging/discharging power at a station exceeds the upper-limit value determined by the SOC status, the excess (or residual) power is assigned to another station on the feeder close to the bank. 

Here, we point out that the above procedure does not necessarily lead to the perfect provision of AS demanded as a regulation signal to DSO. 
Thus, the calculated charging/discharging power at each charging station is refined to achieve the perfect provision. 
This is conducted from the charging station closest to the bank in order. 
By doing this, it would happen that the value of $\sum_{j\in\mathcal{I}_x}(GP_j + BQ_j)$ becomes far from zero. 
Thus, we determine the value of reactive power supplied by EVs such that $\sum_{j\in\mathcal{I}_x}(GP_j + BQ_j)$ again approaches a value around zero. 
For the compensation of reactive power, there exist authorized grid codes for the lower limit of power factor: $0.95$ in Ireland and U.K.\cite{Keane}; $0.9$ in Japan\cite{Tanaka}. 
In this paper, according to the Japanese grid code, we set the upper limit of the charging/discharging (active) power for every charging station to $90\,\%$ of the original upper limit in order to enable us to supply reactive power by all the charging stations. 

\subsection{Proposed Algorithm} 
\label{sec:algorithm}

Suppose that multiple stations and loads are connected to the single feeder grid as shown in Figure~\ref{fig:feeder1}. 
Before algorithm development, we summarize the input data to the algorithm as follows:
\begin{itemize}
\item $P\sub{ref}$: Active power (or value of regulation signal) demanded as AS to DSO that manages the feeder and commands EVs at charging stations to charge, discharge, or stop; 
\item $N\sub{sta}$: Total number of the charging stations;
\item $\xi_{\mathrm{sta},i}\in\{\xi_1,\ldots,\xi_N\}$ ($i=1,\ldots,N\sub{sta}$): Location of $i$-th charging station;
\item $[\underline{P}_i,\overline{P}_i]$: Range of possible charging/discharging power by a group of EVs (in-vehicle batteries) connected to the $i$-th station, where $\underline{P}_i\leq 0$ for charging and $\overline{P}_i\geq 0$ for discharging; 
\item $N\sub{L}$: Total number of loads;
\item $\xi_{\mathrm{L}j}\in\{\xi_1,\ldots,\xi_N\}$ ($j=1,\ldots,N\sub{L}$): Location of $j$-th load; 
\item $P_{\mathrm{L}j}\,(\leq 0)$ ($j=1,\ldots,N\sub{L}$): Power consumption of $j$-th load; and 
\item $G$ and $B$: Feeder's conductance and susceptance per unit-length. 
\end{itemize}
Here, we have $N\sub{sta}+N\sub{L}=N$ and $\{\xi_{\mathrm{sta},1},\ldots,\xi_{\mathrm{sta},N\sub{sta}}\}\cup\{\xi_{\mathrm{L}1},\ldots,\xi_{\mathrm{L},N\sub{L}}\}=\{\xi_1,\ldots,\xi_N\}$, and we note that the indexes $i,j$ of $\xi_{\mathrm{sta},i}$ and $\xi_{\mathrm{L}j}$ are chosen from the end point of the feeder to the starting one; that is, $\xi_{\mathrm{sta},N\sub{sta}}$ and $\xi_{\mathrm{L},N\sub{L}}$ are the nearest station and load to the bank along the feeder line.  
We assume that the above data on in-vehicle batteries are available in the EV-sharing system \cite{Kawashima:CCTA17}, and that the data on loads are available in a DSO. 
Since we intended the primary frequency control reserve, $P\sub{ref}$ is regarded as a part of the amount of active power required for stabilization of grid frequency. 
Several methods for determining the value $P\sub{ref}$ are reported in \cite{Ota,Masuta}. 

\begin{algorithm}[t]
\caption{%
Synthesis of charging/discharging pattern (active and reactive power) of in-vehicle batteries for provision of AS and mitigation of voltage impact
}%
\label{algo1}
\begin{algorithmic}[1]
\Function{main}{}
\State $\{P_{\mathrm{EVs},i}\}_{i=1,\ldots,N\sub{\color{black}sta}} \leftarrow {\rm ACTIVE}(P\sub{ref}) $
\State $\{Q_{\mathrm{EVs},i}\}_{i=1,\ldots,N\sub{\color{black}sta}} \leftarrow {\rm REACTIVE}(\{P_{\mathrm{EVs},i}\}_{i=1,\ldots,N\sub{\color{black}sta}}) $
\EndFunction
\end{algorithmic}
\end{algorithm}
\begin{algorithm}[t]
\caption{%
Determination of charging/discharging (active) power 
}%
\label{algo2}
\begin{algorithmic}[1]
\Function{ACTIVE}{$P\sub{ref}$}
\State $i \leftarrow 1, j \leftarrow 1, P\sub{residual} \leftarrow 0$
\While{$i \leq N\sub{sta}$}
\If{$P\sub{residual} \neq 0$}
\State $P_{\mathrm{EVs},i} \leftarrow P\sub{residual}$
\State $P\sub{residual} \leftarrow 0$
\Else
\State $P_{\mathrm{EVs},i} \leftarrow 0$
\EndIf
\While{$\xi_{\mathrm{L}j}>\xi_{\mathrm{sta},i}$}
\State $P_{\mathrm{EVs},i} \leftarrow P_{\mathrm{EVs},i} - P_{\mathrm{L}j}$
\If{$P_{\mathrm{EVs},i}>\overline{P}_i$}
\State $P\sub{residual} \leftarrow P_{\mathrm{EVs},i} - \overline{P}_i$
\State $P_{\mathrm{EVs},i} \leftarrow \overline{P}_i$
\State $j \leftarrow j + 1$
\State break
\ElsIf{$P_{\mathrm{EVs},i}<\underline{P}_i$}
\State $P\sub{residual} \leftarrow P_{\mathrm{EVs},i} - \underline{P}_i$
\State $P_{\mathrm{EVs},i} \leftarrow \underline{P}_i$
\State $j \leftarrow j + 1$
\State break
\EndIf
\State $j \leftarrow j +1$ 
\EndWhile
\State $P\sub{ref} \leftarrow P\sub{ref} - P_{\mathrm{EVs},i}$
\State $i \leftarrow i + 1$
\EndWhile
\State $i \leftarrow i-1$
\While{$(P\sub{ref} \neq 0) \land (i \geq 1)$}
\If{$P_{\mathrm{EVs},i} + P\sub{ref} > \overline{P}_i$}
\State $P\sub{ref} \leftarrow P\sub{ref} - \overline{P}_i + P_{\mathrm{EVs},i}$
\State $P_{\mathrm{EVs},i} \leftarrow \overline{P}_i$
\State $i \leftarrow i - 1$
\ElsIf{$P_{\mathrm{EVs},i} + P\sub{ref} < \underline{P}_i$}
\State $P\sub{ref} \leftarrow P\sub{ref} - \underline{P}_i + P_{\mathrm{EVs},i}$
\State $P_{\mathrm{EVs},i} \leftarrow \underline{P}_i$
\State $i \leftarrow i - 1$
\Else
\State $P_{\mathrm{EVs},i} \leftarrow P_{\mathrm{EVs},i} + P\sub{ref}$
\State $P\sub{ref} \leftarrow 0$
\EndIf
\EndWhile
\State return $P_{\mathrm{EVs},i}$
\EndFunction
\end{algorithmic}
\end{algorithm}
\begin{algorithm}[t]
\caption{%
Determination of reactive power compensation
}%
\label{algo3}
\begin{algorithmic}[1]
\Function{REACTIVE}{$\{P_{{\rm EVs},i}\}_{i=1,\ldots,N}$}
\State $i \leftarrow 1, j \leftarrow 1, Q\sub{residual} \leftarrow 0$
\While{$i \leq N\sub{sta}$}
\If{$Q\sub{residual} \neq 0$}
\State $Q_{\mathrm{EVs},i} \leftarrow Q\sub{residual}$
\State $Q\sub{residual} \leftarrow 0$
\Else
\State $Q_{\mathrm{EVs},i} \leftarrow 0$
\EndIf
\State $\overline{Q}_i \leftarrow \sqrt{(P_{\mathrm{EVs},i}/0.9)^2 - P_{\mathrm{EVs},i}^2}$
\State $\underline{Q}_i \leftarrow -\sqrt{(P_{\mathrm{EVs},i}/0.9)^2 - P_{\mathrm{EVs},i}^2}$
\While{$\xi_{\mathrm{L}j}>\xi_{\mathrm{sta},i}$}
\State $Q_{\mathrm{EVs},i} \leftarrow GB^{-1}(P_{\mathrm{EVs},i} - P_{\mathrm{L}j})$
\If{$Q_{\mathrm{EVs},i}>\overline{Q}_i$}
\State $Q\sub{residual} \leftarrow Q_{\mathrm{EVs},i} - \overline{Q}_i$
\State $Q_{\mathrm{EVs},i} \leftarrow \overline{Q}_i$
\State $j \leftarrow j + 1$
\State break
\ElsIf{$Q_{\mathrm{EVs},i}<\underline{Q}_i$}
\State $Q\sub{residual} \leftarrow Q_{\mathrm{EVs},i} - \underline{Q}_i$
\State $Q_{\mathrm{EVs},i} \leftarrow \underline{Q}_i$
\State $j \leftarrow j + 1$
\State break
\EndIf
\State $j \leftarrow j +1$ 
\EndWhile
\State $i \leftarrow i + 1$
\EndWhile
\State return $Q_{\mathrm{EVs},i}$
\EndFunction
\end{algorithmic}
\end{algorithm}

Thus, in order to provide the AS demand power and reduce the deviation of distribution voltage, we now propose an algorithm to determine the amount of charging/discharging (active) power by a group of EVs at $i$-th station, $P_{\mathrm{EVs},i}\in[\underline{P}_i,\overline{P}_i]$ as well as supplying reactive power, $Q_{\mathrm{EVs},i}$. 
This procedure is summarized in Algorithms~1, 2, and 3. 
Algorithm~1 consists of Algorithms~2 and 3 that determine $P_{\mathrm{EVs},i}$ and $Q_{\mathrm{EVs},i}$, respectively. 

Algorithm~2 describes the procedure of determining the charging/discharging power $P_{\mathrm{EVs},i}$ for the given regulation signal $P\sub{ref}$ and is divided into the two parts. 
Firstly, from the end point of the feeder, the value of $P_{\mathrm{EVs},i}$ at $i$-th station is calculated in a manner such that the sum $\sum_{i\in\mathcal{I}_x}(GP_i + BQ_i)$ under $Q_i=0$ approaches a value around zero. 
Hence, the deviation of distribution voltage from the nominal value is reduced. 
Here, the range $[\underline{P}_i,\overline{P}_i]$ of possible charging/discharging power depends on the number of EVs connected there, SOC of the corresponding batteries, inverter capacity of the battery chargers, plan and request of use of shared EVs, and so on. 
The residual of power happens when the calculated power for $i$-th station is greater than $\overline{P}_i$ (see line~12 in Algorithm~2) or smaller than $\underline{P}_i$ (see line~17 in Algorithm~2). 
This residual is incorporated for calculation of the next station, $P_{\mathrm{EVs},i+1}$ (see lines~4--6 in Algorithm~2). 
Secondly, from the charging station closest to the bank, $P_{\mathrm{EVs},i}$ is refined in a manner such that the total sum of $P_{\mathrm{EVs},i}$ is equal to the given regulation signal $P\sub{ref}$ (see lines~29--42 in Algorithm~2). 

Algorithm~3 follows Algorithm~2 and describes the procedure of determining the supplying reactive power $Q_{\mathrm{EVs},i}$ for supporting the voltage regulation above. 
The calculated $P_{\mathrm{EVs},i}$ may not be enough to setting $\sum_{i\in\mathcal{I}_x}(GP_i + BQ_i)$ close to zero due to the value of regulation signal and the existing range $[\underline{P}_i,\overline{P}_i]$. 
In Algorithm~3, the value of $Q_{\mathrm{EVs},i}$ at $i$-th station is calculated in a manner such that the sum $\sum_{i\in\mathcal{I}_x}(GP_i + BQ_i)$ approaches a value around zero. 
Thus, the supply of reactive power can compensate the loss of regulation effort by charging/discharging power $P_{\mathrm{EVs},i}$. 
In this paper, according to the Japanese grid code \cite{Tanaka}, we suppose that the feasible range of power factor is set at $[0.9,1]$, and associated range $[\underline{Q}_i,\overline{Q}_i]$ of supplying reactive power is set with the pre-calculated $P_{\mathrm{EVs},i}$ (see lines~10 and 11 in Algorithm~3). 
In the above calculation of $Q_{\mathrm{EVs},i}$, the residual of reactive power happens if the calculated reactive power is greater than $\overline{Q}_i$ (see line~14 in Algorithm~3) or smaller than $\underline{Q}_i$ (see line~19 in Algorithm~3). 
This residual is incorporated for determination of the next station, $Q_{\mathrm{EVs},i+1}$ (see lines~4--6 of Algorithm~3) in the same manner as $P_{\mathrm{EVs},i+1}$. 

Note that grid-connected inverters are conventionally initiated to provide reactive power when their regulation of active-power output is insufficient for voltage control \cite{Keane}.  
A flow chart of initiating the supply of reactive power is decided in the grid code.  
In this paper, to evaluate the effectiveness of synthesized pattern of active and reactive power, we simply assume that any inverter in this paper can provide both active and reactive power in any situation.

\section{Numerical Experiments}
\label{sec:demo}

\subsection{Single Feeder}
\label{subsec:single}

\begin{figure}[t]
\centering
\includegraphics[width=.65\hsize]{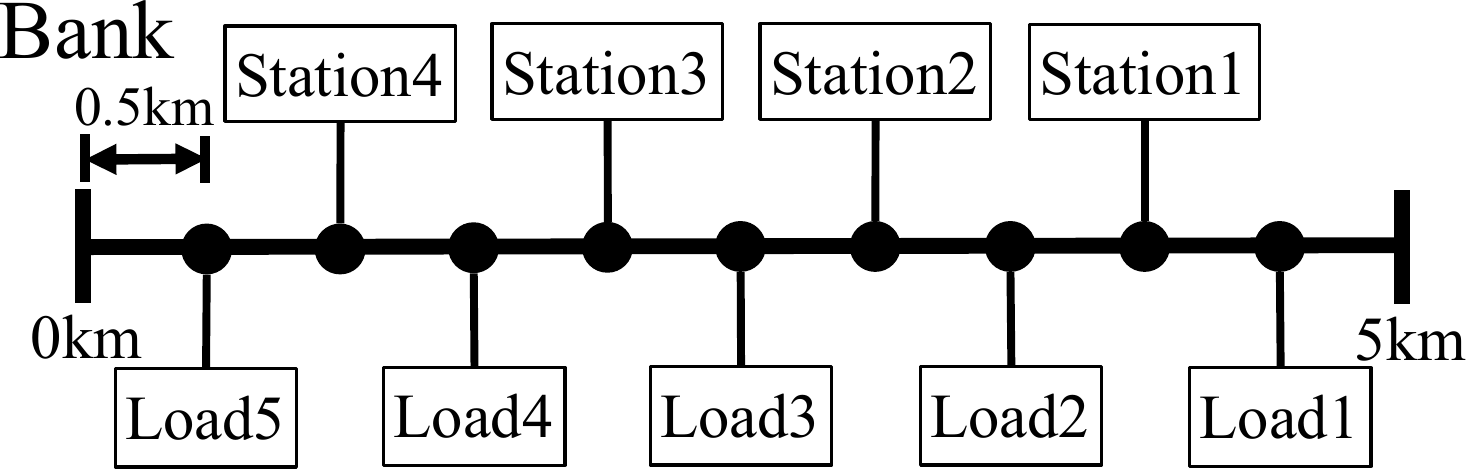}
\caption{%
Single feeder model. 
The feeder has $4$ charging stations and $5$ loads located at a common interval ($0.5\,\U{km}$). 
The starting and end points are assumed to have no load for simplicity of the current study. 
}%
\label{fig:feeder3}
\end{figure}

\begin{figure*}[t]
\centering
\begin{minipage}{.495\textwidth}
\centering
\includegraphics[width=\hsize]{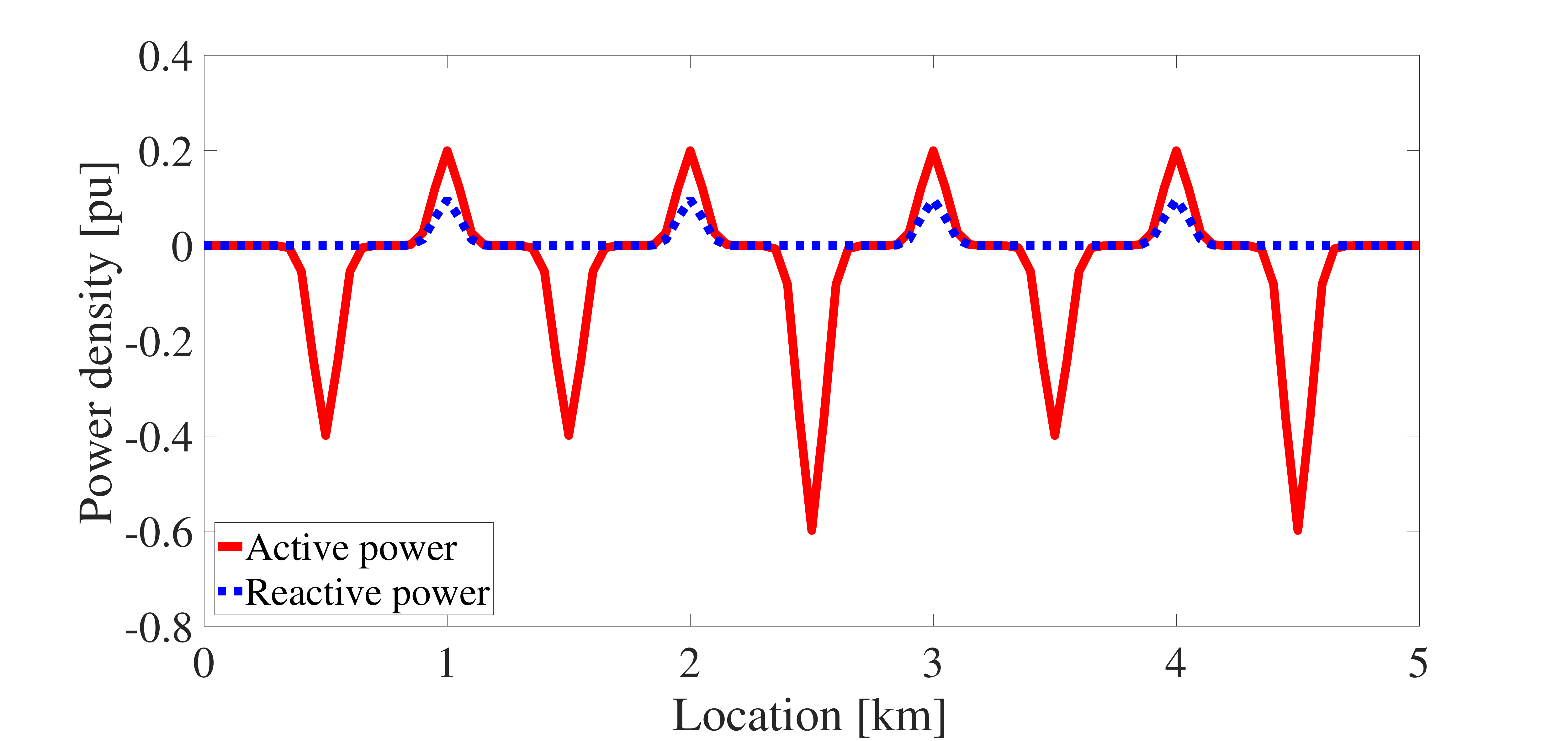}
\includegraphics[width=\hsize]{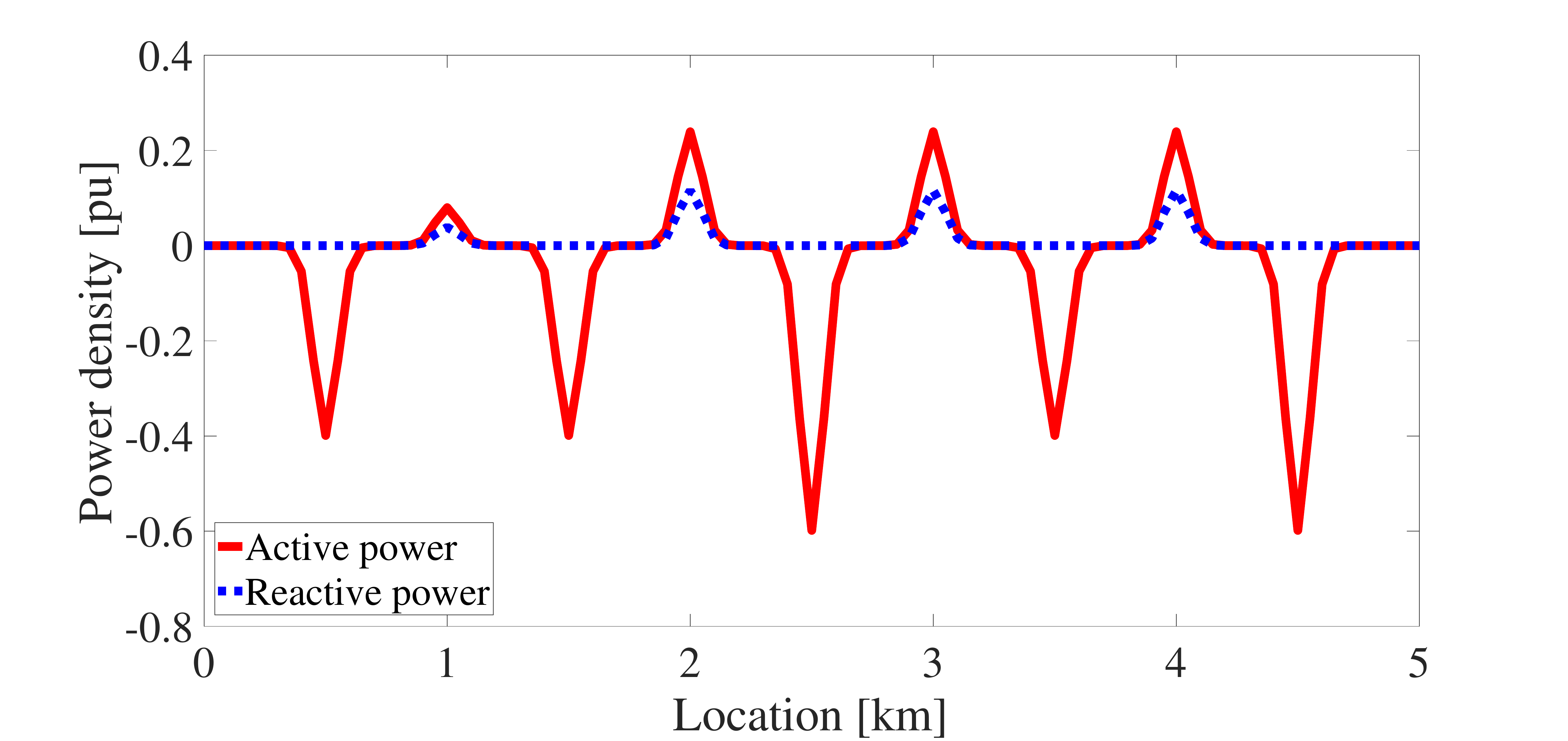}
\subcaption{%
Spatial patterns of active and reactive power
}%
\label{fig:patterns_4}
\end{minipage}
\begin{minipage}{.495\textwidth}
\centering
\includegraphics[width=\hsize]{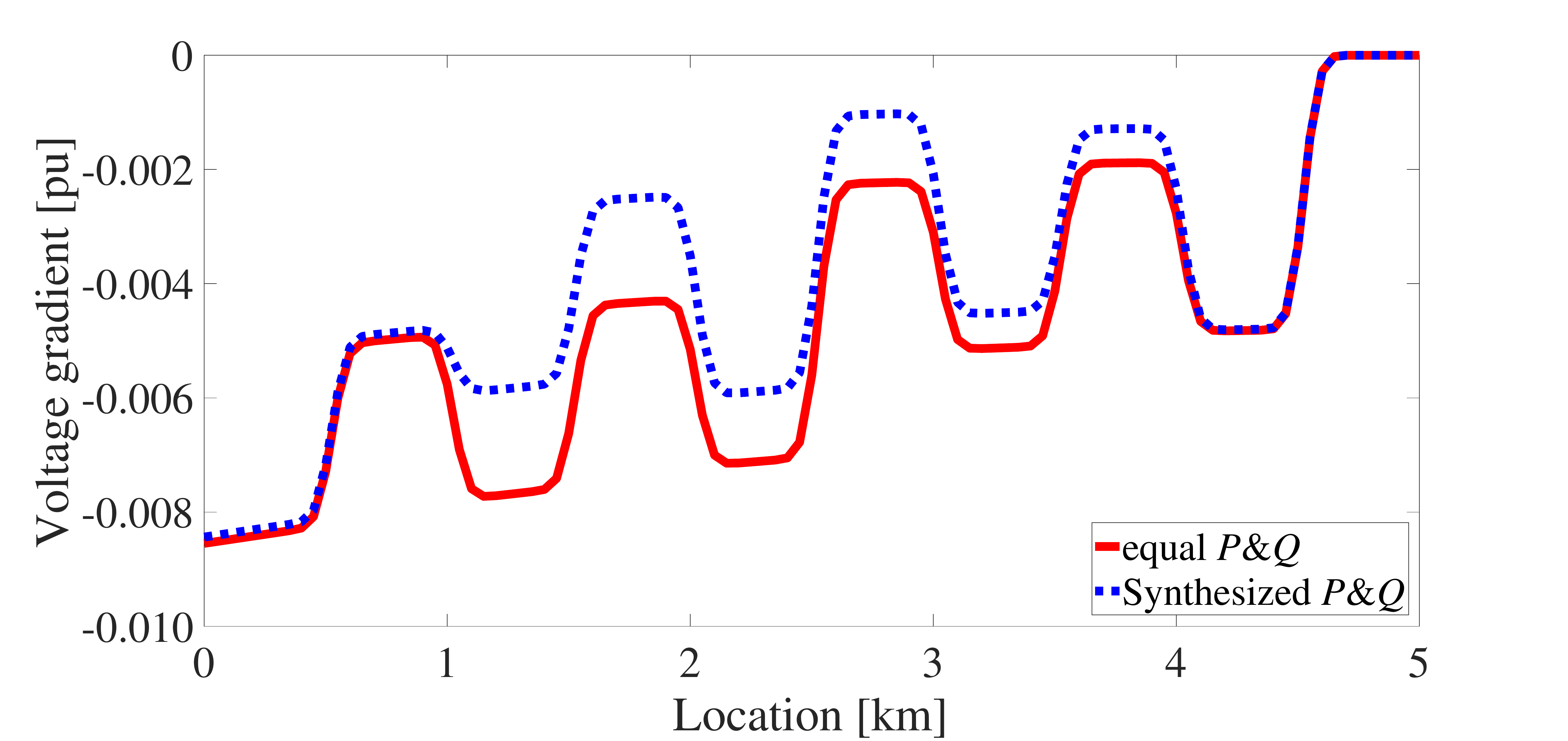}
\includegraphics[width=\hsize]{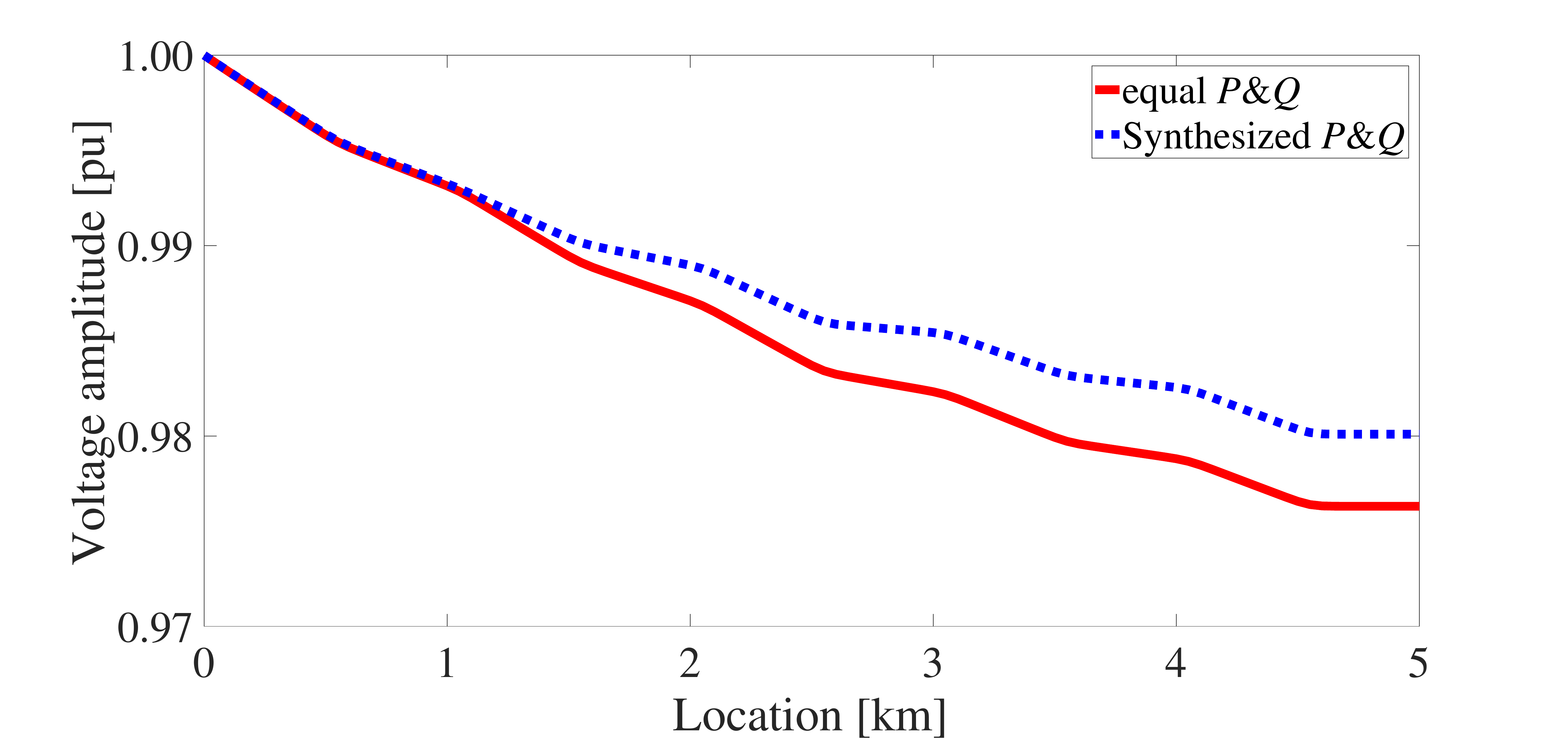}
\subcaption{%
Spatial profiles of voltage gradient and amplitude
}%
\label{fig:simulation_4}
\end{minipage}
\vspace*{1mm}
\caption{%
Results on the proposed algorithm for synthesizing the spatial pattern of discharging operations of in-vehicle batteries for the single feeder. 
In (a) the \emph{red,\,solid} lines represent spatial patterns of active power and the \emph{blue,\,dotted} lines do those of reactive power. 
The top of the figure (a) shows the consumption of loads (negativeness) and the uniform pattern of discharging power (positiveness) and of supplying reactive power (positiveness, too). 
The bottom of the figure (a) shows the consumption of loads and the synthesized pattern of discharging power and supplying reactive power based on the proposed algorithm. 
The figure (b) shows the numerical simulations of the nonlinear ODE (\ref{eqn:ode}) incorporating with the power density functions based on the figure (a). 
The \emph{red,\,solid} lines denote the simulation results for the uniform pattern (on the top of the figure (a)) and the \emph{blue,\,dashed} lines for the synthesized pattern (on the bottom of the figure (a)). 
}%
\label{fig:demo1}
\end{figure*}

In this sub-section, we evaluate the effectiveness of the proposed algorithm with direct numerical simulations of (\ref{eqn:ode}) for the single feeder model in Figure~\ref{fig:feeder3}. 
This feeder has loads and charging stations located at a common interval ($0.5\,\U{km}$). 
We have assumed that no load exists at the end point of the feeder. 
The secondary voltage at the bank is regulated at $6.6\,\U{kV}$. 
The loading capacity of the bank transformer is set at $12\,\U{MVA}$, and the feeder's resistance (or reactance) at $0.227\,\U{\Omega/km}$ (or $0.401\,\U{\Omega/km}$). 
These values are based on the standard setting of medium-voltage distribution grids in Japan{\color{black}; the loading capacity ($12\,\U{MVA}$) corresponds to the base power in this subsection.  Thereby,} 
we use per-unit system for numerical simulations of the single feeder model. 
The conductance $G$ (or susceptance $B$) per unit-length is calculated as $3.881$ (or $6.856$) in per-unit system (hence $G/B$ is about $5.661\times 10^{-1}$). 
The total amount of the loads is set as $30\,\%$ of the loading capacity of the bank. 
The loads are represented as constant power model as stated above. 
We suppose that each EV has rated power output of $4\,\U{kW}$ ($3.3\times 10^{-4}$) \cite{Clement1}, and each station has the limit of maximum 100 EVs for simultaneous charging and discharging operations, implying $\underline{P}_i=-3.3\times 10^{-2}$ and $\overline{P}_i=3.3\times 10^{-2}$. 
The regulation signal $P\sub{ref}$ to DSO managing the single feeder with charging stations is set at $1.00\times10^{-1}$, implying 10\% of the loading capacity of the bank. 

For comparison, we consider the two cases on synthesized patterns of charging/discharging (active) power and supplying reactive power. 
One case is based on the proposed algorithm, and the resulting patterns on active and reactive power are normally nonuniform for every charging station. 
The other case considers an uniform pattern of charging/discharging power, which is equally allocated and is derived by dividing the regulation signal $P\sub{ref}$ by the number of charging stations \cite{Shimizu,Masuta}. 
Also, in the second case, we consider an uniform pattern of the supplying reactive power for every station, in which we use the common value of charging/discharging power multiplied by a constant power factor, in this paper, $0.9$ \cite{Tanaka}.

In order to avoid the difficulty of numerical simulations of (\ref{eqn:ode}) due to the delta function in $p(x)$ of (\ref{eqn:pq}), as in \cite{Susuki2} we use the following coarse-graining of the function with Gaussian function:
\begin{equation}
p(x)\sim\sum_{i=1}^{N}\frac{P_i}{\sqrt{2\pi\sigma^2}}
\exp\!\left(-\frac{(x-\xi_i)^2}{2\sigma^2}\right)
\end{equation}
where we regard $x,\xi_i$ as scalars, and $\sigma$ is the standard deviation. 
$\sigma$ was fixed at $0.05\,\U{km}$, which was much smaller than the interval ($0.5\,\U{km}$). 
The reactive-power $q(x)$ is also treated in the same manner as $p(x)$ above. 
This coarse-graining will be used for plotting the functions and spatial patterns of active and reactive power. 

Figure~\ref{fig:patterns_4} shows the spatial patterns of active and reactive power that result in the power density functions $p(x)$ and $q(x)$. 
The \emph{red,\,solid} lines denote active power and the \emph{blue,\,dotted} lines do reactive power. 
The top of the figure (a) shows the consumption of the 5 loads (negativeness) and the uniform discharging power (positiveness) from the 4 stations, which is equally allocated, that is, derived by dividing the regulation signal $P\sub{ref}$ by the number of charging stations. 
The uniform supplying reactive power (positiveness, that is, leading-phase) is also shown in the same figure. 
The bottom of the figure (a) shows the consumption of loads and the synthesized pattern of discharging power and supply reactive power based on the proposed algorithm. 
The total sum of the discharging power at the 4 stations is equal to $P\sub{ref}=1.00\times10^{-1}$ for both the cases, as shown in Tables~\ref{tab:pq_single_equal} and \ref{tab:pq_single_synth}. 
This implies that the provision of active power from EVs is realized with the proposed algorithm.

\begin{table}[t]
\centering
\scriptsize
\caption{%
Values of spatial patterns of the $4$ charging stations in the single feeder model
}%
\subcaption{Uniform case (on the top of Figure~3a)}
\label{tab:pq_single_equal}
\begin{tabular}{c||c|c|c|c|c} \hline
&St.\,1&St.\,2&St.\,3&St.\,4 & Total\\ \hline \hline
$P_i / 10^{-2}$&$2.50$&$2.50$&$2.50$&$2.50$ & $\bf 10.0$\\ \hline
$Q_i / 10^{-2}$&$1.21$&$1.21$&$1.21$&$1.21$ & -- \\ \hline
\end{tabular}
\vspace*{5mm}
\subcaption{Proposed case (on the bottom of Figure~3a)}
\label{tab:pq_single_synth}
\begin{tabular}{c||c|c|c|c|c} \hline
&St.\,1&St.\,2&St.\,3&St.\,4 & Total \\ \hline \hline
$P_i / 10^{-2}$&$1.00$&$3.00$&$3.00$&$3.00$ & $\bf 10.0$\\ \hline
$Q_i / 10^{-2}$&$0.48$&$1.45$&$1.45$&$1.45$ & --\\ \hline
\end{tabular}
\end{table}

Figure~\ref{fig:simulation_4} shows the numerical solutions of the nonlinear ODE (\ref{eqn:ode}) incorporating with the power density functions based on Figure~\ref{fig:patterns_4}. 
The top of the figure (b) shows the voltage gradient for the two cases, which are the uniform pattern (on the top of Figure~\ref{fig:patterns_4}) denoted by \emph{red,\,solid} lines and the synthesized pattern (on the bottom of Figure~\ref{fig:patterns_4}) by \emph{blue,\,dashed} lines. 
The associated voltage amplitude is shown in the bottom of the figure (b). 
The top of the figure (b) shows that the loads and charging stations are clearly located by detecting the change points of voltage gradient by comparison with voltage amplitude in the bottom figure. 
By comparison of the two cases, we see that the voltage gradient becomes close to zero in the synthesized pattern, and hence the deviation of voltage amplitude from nominal is decreased. 
This clearly shows that the proposed algorithm is effective for the mitigation of voltage impact.

The voltage amplitude in the bottom of Figure~\ref{fig:simulation_4} is the result of synthesized pattern of discharging operations in Table~\ref{tab:pq_single_synth}.  
based on Algorithms 1 to 3.  
Since the algorithms do not use any optimization, the synthesized pattern is not necessarily optimal in a suitable sense such as the $\mathcal{L}_2$ norm $\int_0^{L} \{v(x)-1\}^2dx$ or $\int_0^{L}\{w(x)\}^2dx$.   
An iterative calculation is required for finding the optimal pattern of spatial charging/{\color{black}discharging}.  
We here contend that the proposed algorithm without iteration is significant to the provision of primary frequency control reserve because its execution is terminated quickly. 
{\color{black}Indeed}, 
for a single feeder, only three steps are executed for determining the pattern: Construction of power density function based on a load profile, (Algorithm 2) calculation of active power to achieve $P\sub{ref}$, and (Algorithm 3) calculation of reactive power.

\subsection{Multiple feeders}
\label{subsec:multi}

\begin{figure}[t]
\centering
\includegraphics[width=.65\hsize]{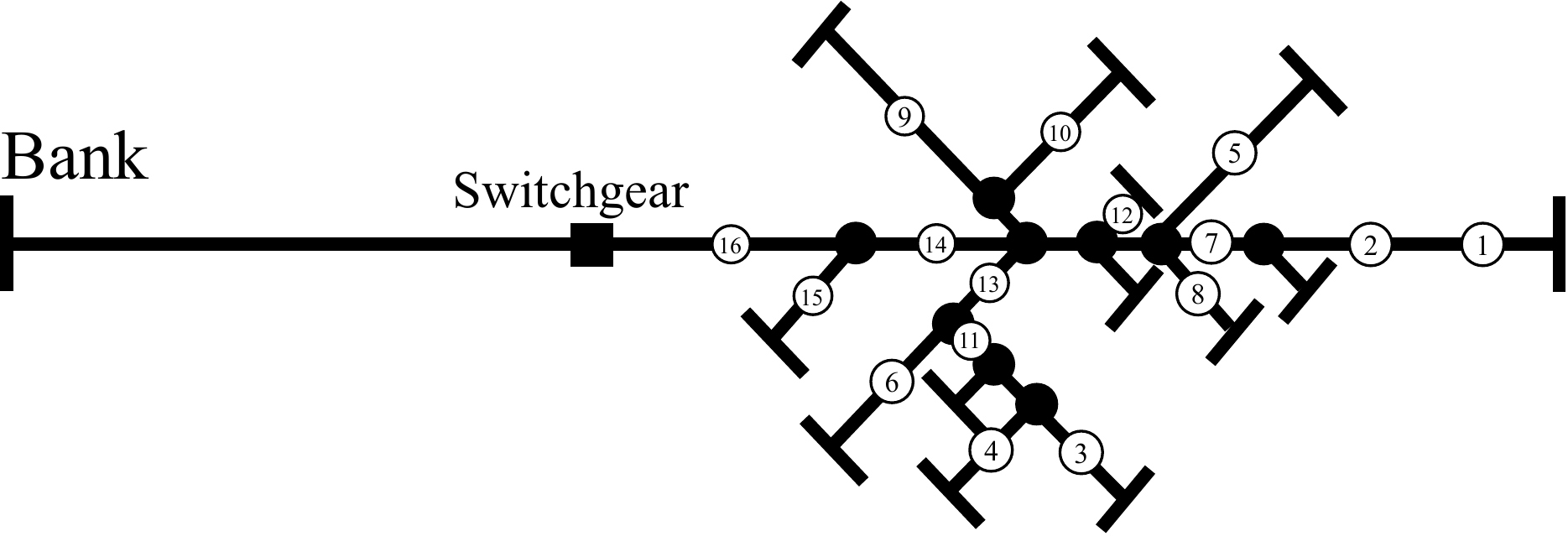}
\caption{%
Model of multiple feeders based on a practical distribution grid of residential area in Japan. 
The 16 charging stations are virtually installed and denoted by the \emph{circled numbers}. 
}%
\label{fig:feeder4}
\centering
\includegraphics[width=0.65\hsize]{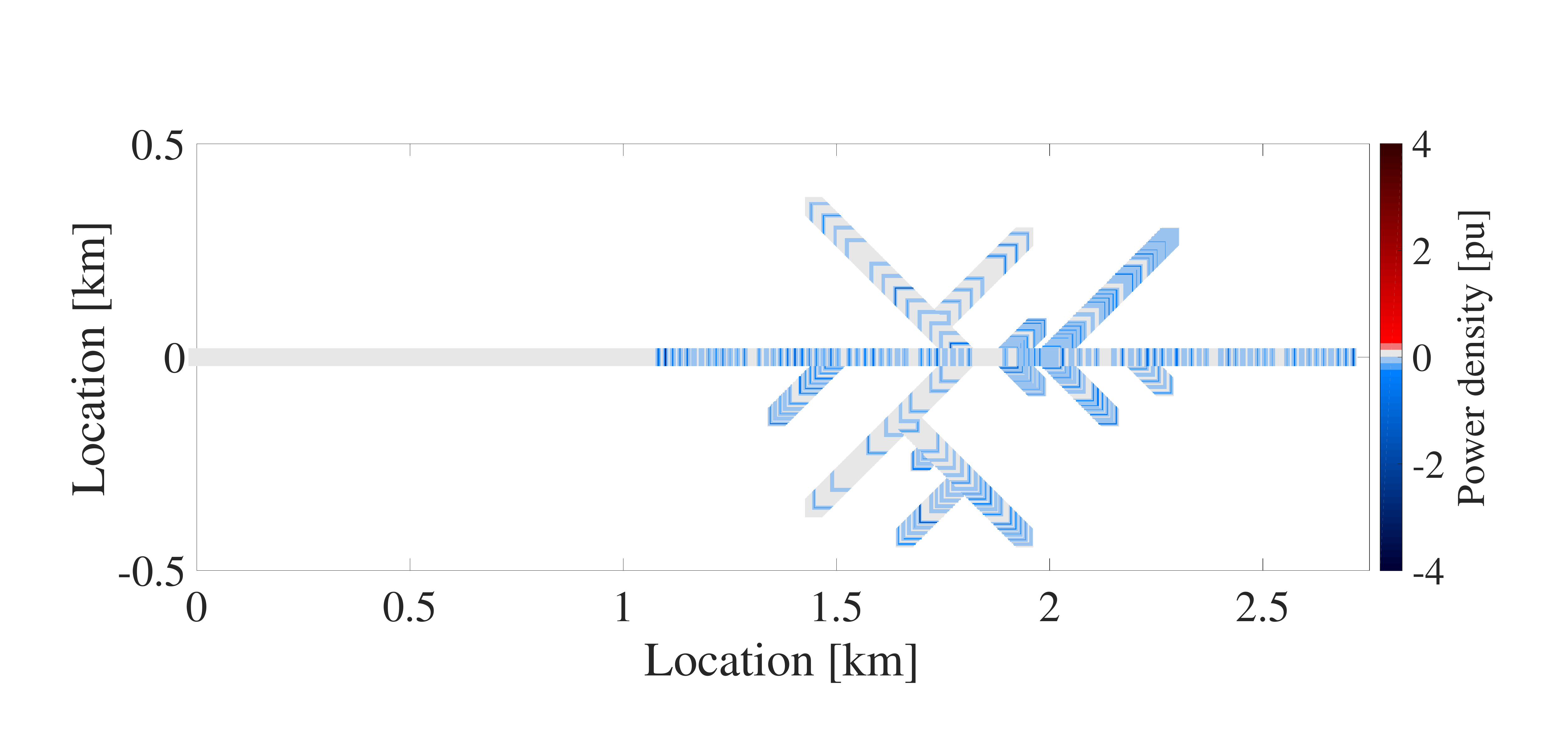}
\caption{%
Spatial distribution of loads in the model of multiple feeders. 
The \emph{black} parts (negative values of density) represent the power consumption by loads via pole transformers. 
}%
\label{fig:load2_19}
\end{figure}

In this sub-section, we evaluate the effectiveness of the proposed algorithm for the multiple feeders shown in Figure~\ref{fig:feeder4}. 
This model is based on a practical distribution grid of residential area in western Japan and is provided by a power company. 
It has multiple bifurcations denoted by \emph{black circles} and 16 charging stations denoted by \emph{circled numbers}. 
The sum of the lengths of all the feeders is $5.75\,\rm{km}$. 
The model has one switchgear at $1\,\rm{km}$ from the bank and has no load connected to the feeder between the switchgear and the bank. 
We assume that the secondary voltage at the bank is regulated at $6.6\,\U{kV}$. 
The loading capacity of the bank is set at $20\,\U{MVA}$. 
The model has $232$ pole transformers distributed along the feeders from the switchgear to each end. 
All the pole transformers are connected to a total of $4587$ residential households. 
The residential loads used below are based on practical measurement at 19 o'clock in summer in Japan.\footnote{The detailed data on the feeders and loads could not be published following an agreement with the collaborators.} 
The corresponding power density function for loads as constant power model is shown in Figure~\ref{fig:load2_19}.  
The function is constructed in the same way as in the single-feeder case.   
In the figure, the positiveness (\emph{black}) implies the discharging operation by batteries, and the negativeness (\emph{black}) does the charging operation or the power consumption by load. 
The \emph{black} part on the feeders therefore represents the locations connected to the residential loads through the pole transformers. 
In addition to the loads, we assume that each station has the limit of maximum 50 EVs for simultaneous charging or discharging, where each EV has the rated power output of $4\,\U{kVA}$ {\color{black}($3.33\times10^{-4}$ in the per unit system of the multiple feeders; its base power corresponds to the loading capacity of the bank, $20\,\U{MVA}$)}. 
This implies the range of charging/discharging power of each station, $\underline{P}_i=-1.00\times 10^{-2}$ 
and $\overline{P}_i=1.00\times 10^{-2}$. 
Also, we set the regulation signal $P\sub{ref}$ to DSO managing the multiple feeders with charging stations at $1.00\times10^{-2}$, implying 1\% of the loading capacity of the bank. 

For comparison, we consider the two cases on synthesized patterns of charging/discharging (active) power and supplying reactive power. 
One case is based on the proposed algorithm, and the resulting patterns on active and reactive power are normally nonuniform for every charging station. 
The other case is the same as in Section~\ref{subsec:single} and considers a uniform pattern of charging/discharging power, which is equally allocated and is derived by dividing the regulation signal $P\sub{ref}$ by the number of charging stations, in this model, $16$. 
Also, in the second case, we consider an uniform pattern of the supplying reactive power for every station, in which we use the common value of charging/discharging power multiplied by a constant power factor, in this paper, $0.9$ \cite{Tanaka}.

Here, we summarize how to apply the proposed algorithm for the single feeder to the multiples ones. 
Firstly, the amount of charging/discharging (active) power is calculated for each feeder based on Algorithm~2. 
That is, the amount of active power is calculated such that $\sum_{j\in\mathcal{I}_x}(GP_j + BQ_j)$ approaches a value around zero as shown in lines~3--27 of Algorithm~2. 
For a single feeder linking two different bifurcations points, if $P_{\rm{residual}}$ is not zero at the bifurcation point close to the bank, then $P_{\rm{residual}}$ is assigned to the charging station that is the closest (along the feeder) from the bifurcation point (if exists) on the feeder linked to the bank. 
Secondly, from the charging station closest to the bank, $P_{\mathrm{EVs},i}$ is refined in a manner such that the total sum of $P_{\mathrm{EVs},i}$ is equal to the given regulation signal $P\sub{ref}$ as shown in lines~29--42 of Algorithm~2.
Finally, the compensation of reactive power is determined for each feeder based on Algorithm~3. 
For a single feeder linking two different bifurcations points, if $Q_{\rm{residual}}$ is not zero at the bifurcation point close to the bank, then $Q_{\rm{residual}}$ is assigned to the charging station on another feeder in the similar manner as in $P_{\rm residual}$. 

\begin{figure*}[t]
\centering
\vspace{-5mm}
\begin{minipage}{0.495\textwidth}
\includegraphics[width=\hsize]{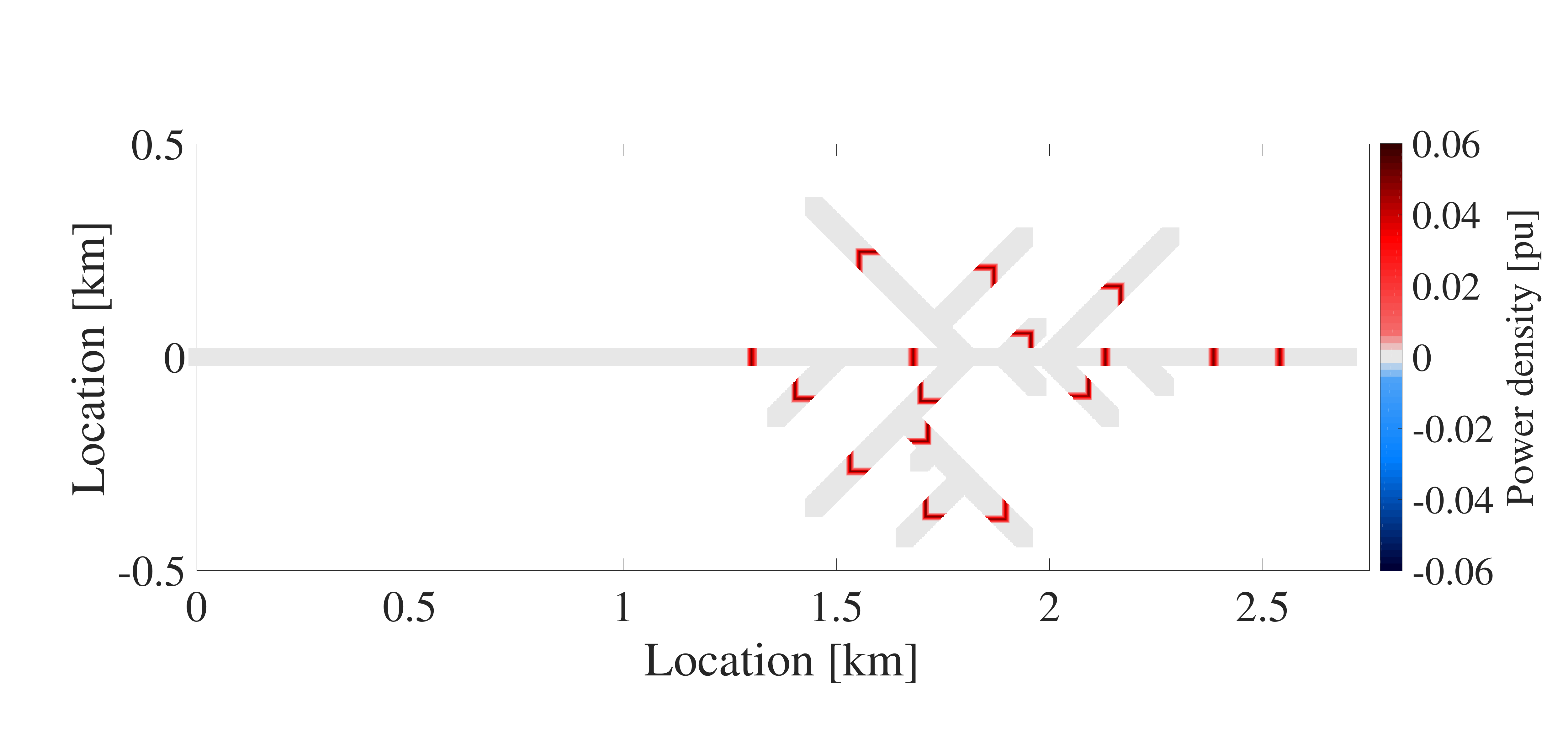}
\vspace{-8mm}
\subcaption{%
\footnotesize
Equally allocated $P$
}%
\label{fig:p_equal}
\end{minipage}
\begin{minipage}{0.495\textwidth}
\includegraphics[width=\hsize]{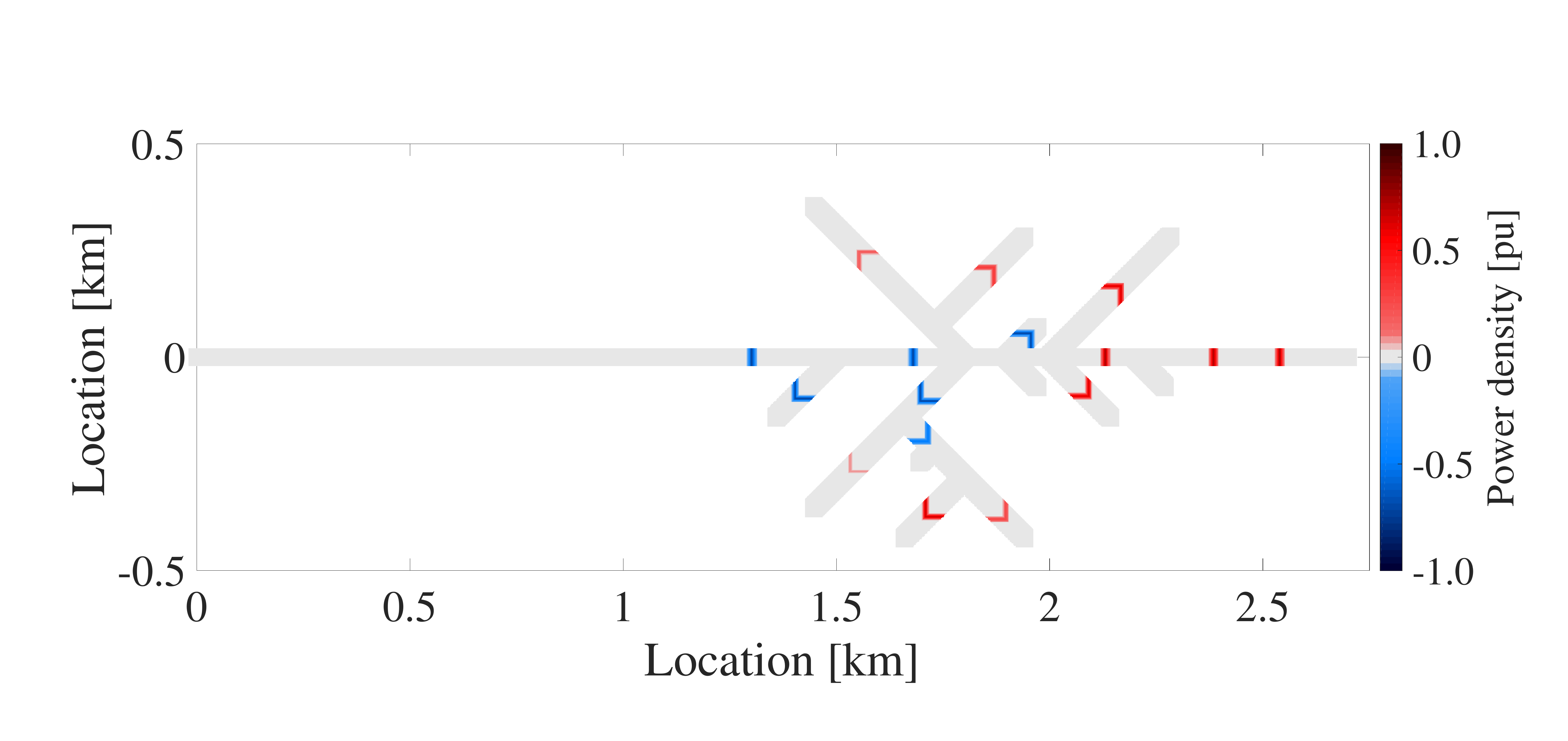}
\vspace{-8mm}
\subcaption{%
\footnotesize
Synthesized $P$
}%
\label{fig:p_synth}
\end{minipage}
\caption{%
Results on charging/discharging patterns at the 16 charging stations. 
The \emph{black} marks represent the discharging operation and the \emph{black} marks the charging operation.
The \emph{left} figure (a) is the result on equally allocated pattern of the given regulation signal, and the \emph{right} figure (b) is the result on the proposed algorithm. 
}%
\label{fig:syn_p_4}
\end{figure*}

Figure~\ref{fig:syn_p_4} shows the results on charging/discharging patterns of active power at the 16 stations. 
The equally allocated pattern is shown in Figure~\ref{fig:p_equal}, and the synthesized pattern based on the proposed algorithm is shown in Figure~\ref{fig:p_synth}. 
Since the synthesis is based on the power density function for loads in Figure~\ref{fig:load2_19} and the regulation signal for AS, we see in the synthesized pattern that multiple stations close to each end of the feeders exhibit the discharging operation (positive value of density). 
As shown below, the discharging operation is assigned such that the deviation of distribution voltage around the ends is reduced. 
In contrast to this, the six stations closed to the bank exhibit the charging operation in order to provide the amount of the regulation signal. 
In fact, the total sum of the amounts of discharging and charging power at the 16 stations is equal to the given regulation signal $P\sub{ref}=1.00\times10^{-2}$ as shown in Tables~\ref{tab:pq_multiple_equal} and \ref{tab:pq_multiple_synth}. 
Therefore, we conclude for the model of multiple feeders that the provision of active power from EVs is realized with the proposed algorithm. 

\begin{figure*}[t]
\centering
\begin{minipage}{0.495\textwidth}
\includegraphics[width=\hsize]{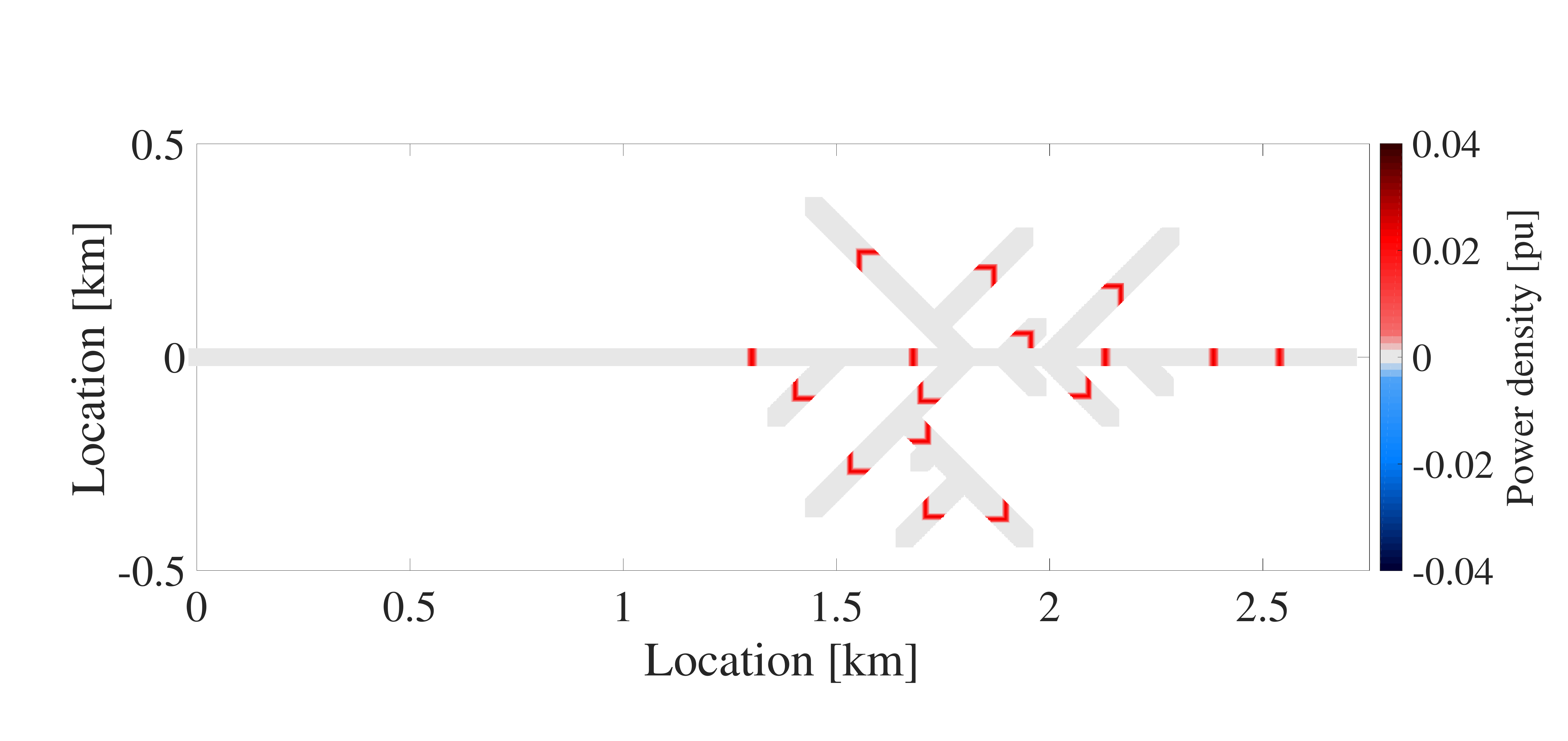}
\vspace{-8mm}
\subcaption{%
\footnotesize
Uniform $Q$ under equally allocated $P$
}%
\label{fig:rea_q_equal}
\end{minipage}
\begin{minipage}{0.495\textwidth}
\includegraphics[width=\hsize]{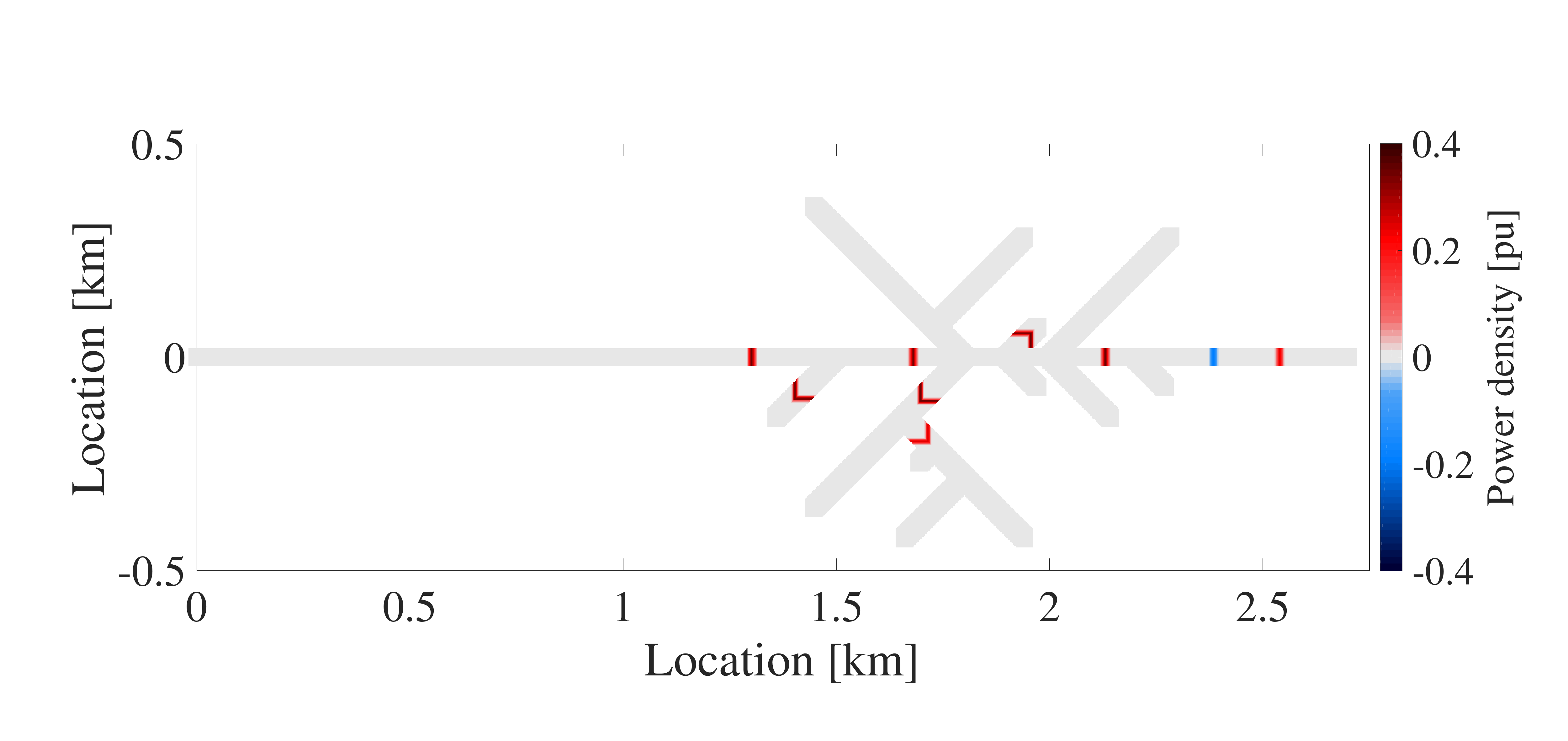}
\vspace{-8mm}
\subcaption{%
\footnotesize
Synthesized $Q$ under synthesized $P$
}%
\label{fig:rea_q_synth}
\end{minipage}
\caption{%
Results on patterns of supply reactive power at the 16 charging stations. 
The \emph{black} marks represent the leading-phase operation and the \emph{black} marks the lagging-phase operation. 
The \emph{left} figure (a) is the result on the constant power-factor control under the equally allocated patten in Figure~\ref{fig:p_equal}. 
The \emph{right} figure (b) is the result on the proposed algorithm under the synthesized pattern in Figure~\ref{fig:p_synth}.
}%
\label{fig:rea_q_4}
\end{figure*}

Figure~\ref{fig:rea_q_4} shows the results on patterns of supplying reactive power. 
In the figure, the positive value (\emph{black}) represents the supply of leading-phase reactive power and the negative value (\emph{black}) does the supply of lagging-phase reactive power. 
The uniform supply under the equally allocated pattern is shown in Figure~\ref{fig:rea_q_equal} where all the reactive power is leading-phase. 
The synthesized result based on the proposed algorithm is shown in Figure~\ref{fig:rea_q_synth}. 
In the synthesized case, since the supply reactive power at each station is determined with the term $\sum_{j\in\mathcal{I}_x}(GP_j + BQ_j)$ in (\ref{eqn:w}), we see the difference of magnitudes of the supplying reactive power (although this is not vivid in Figure~\ref{fig:rea_q_synth}), see Table~\ref{tab:pq_multiple_synth}. 

\begin{figure*}[t]
\centering
\vspace{-5mm}
\begin{minipage}{0.495\textwidth}
\includegraphics[width=\hsize]{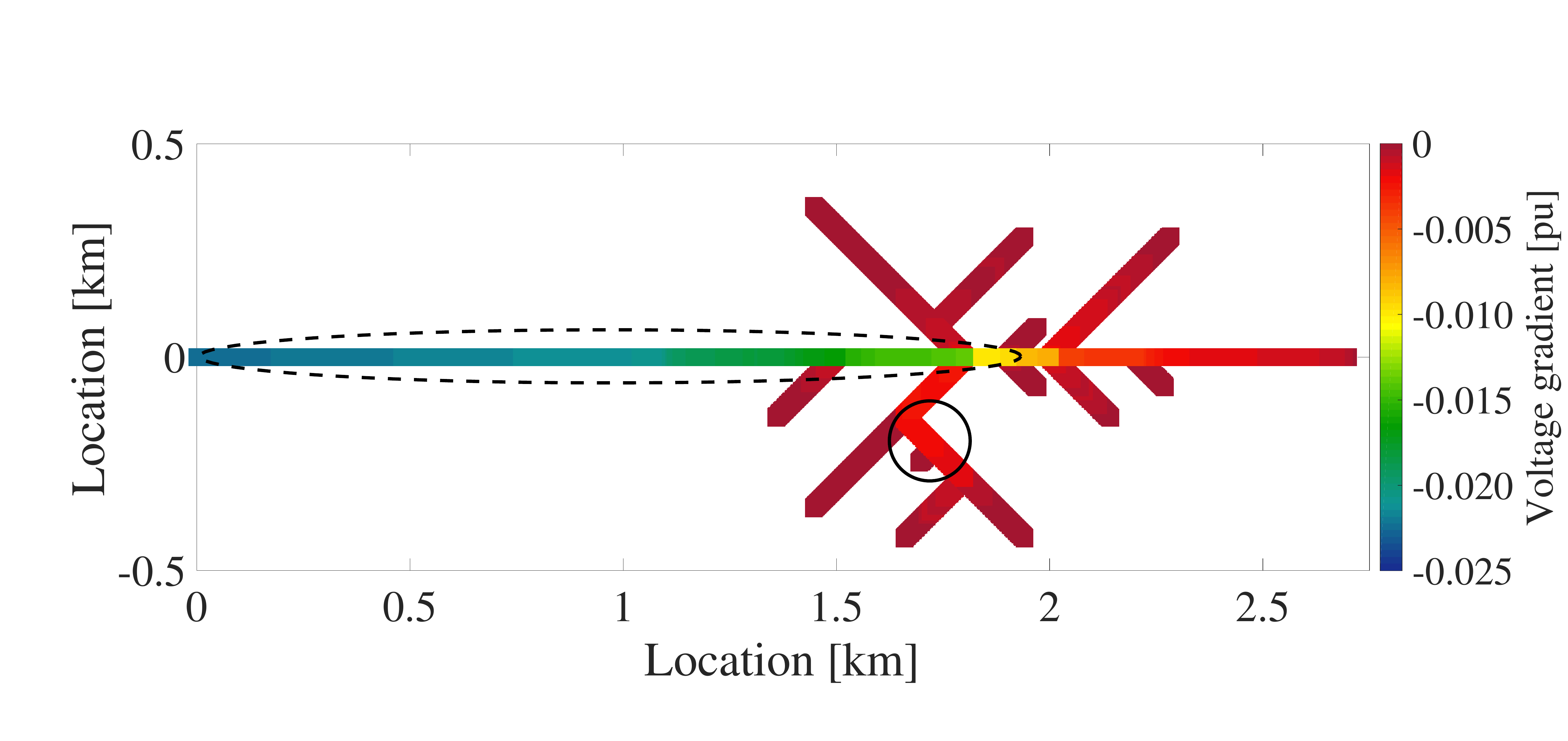}
\vspace{-5mm}
\end{minipage}
\begin{minipage}{0.495\textwidth}
\includegraphics[width=\hsize]{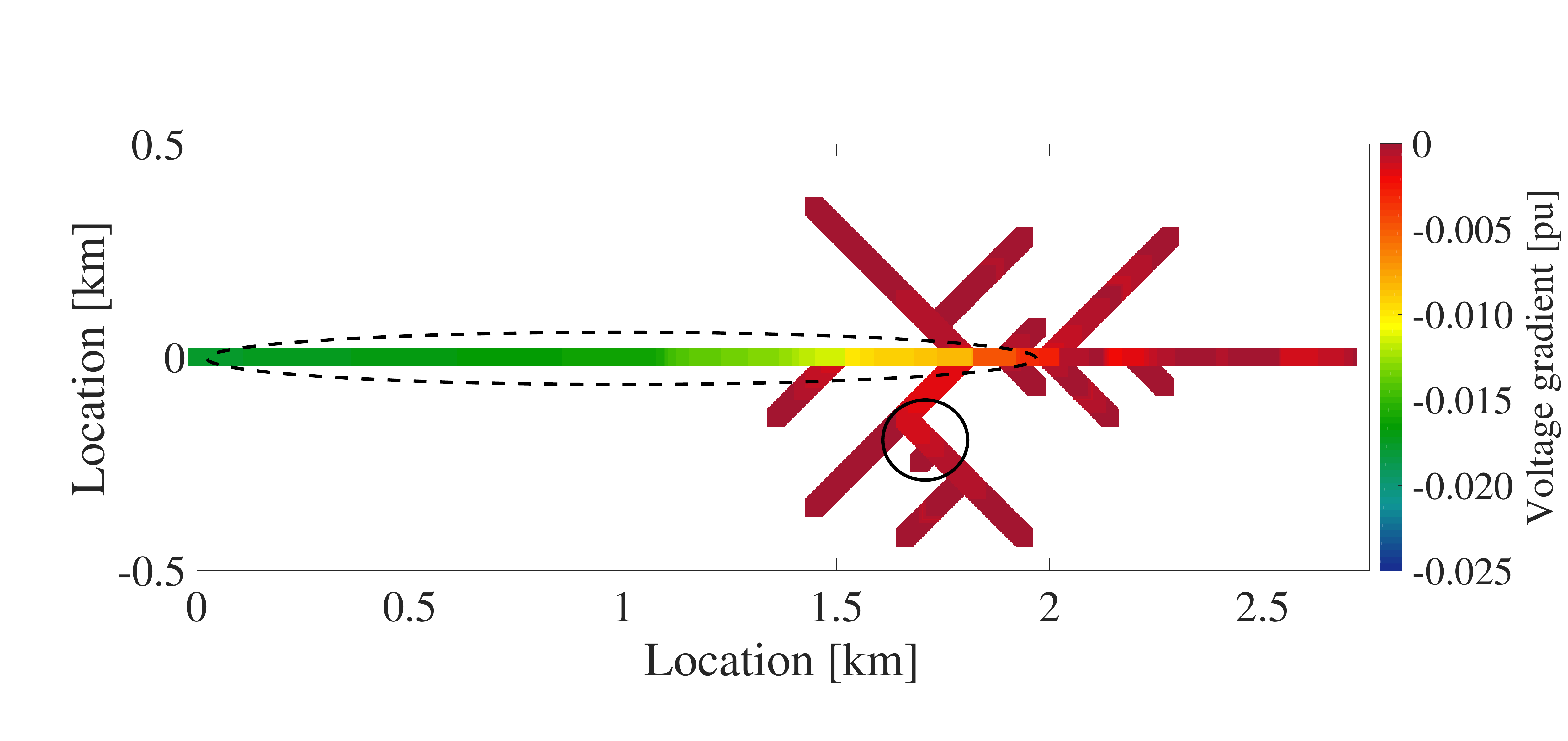}
\vspace{-5mm}
\end{minipage}
\begin{minipage}{0.495\textwidth}
\includegraphics[width=\hsize]{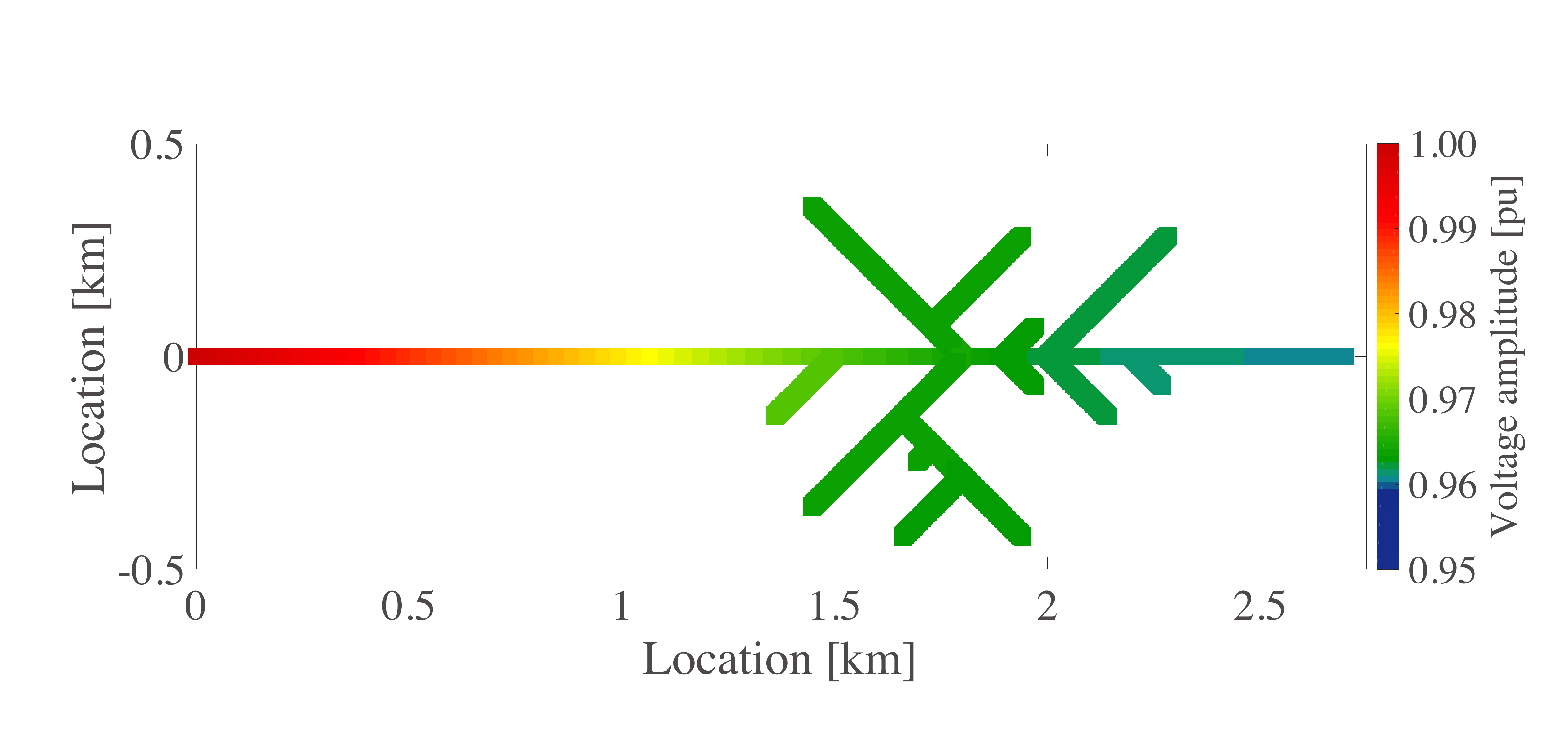}
\vspace{-4mm}
\subcaption{%
\footnotesize
Equally allocated $P$ and uniform $Q$
}%
\label{fig:rea_v_equal-simu}
\end{minipage}
\begin{minipage}{0.495\textwidth}
\includegraphics[width=\hsize]{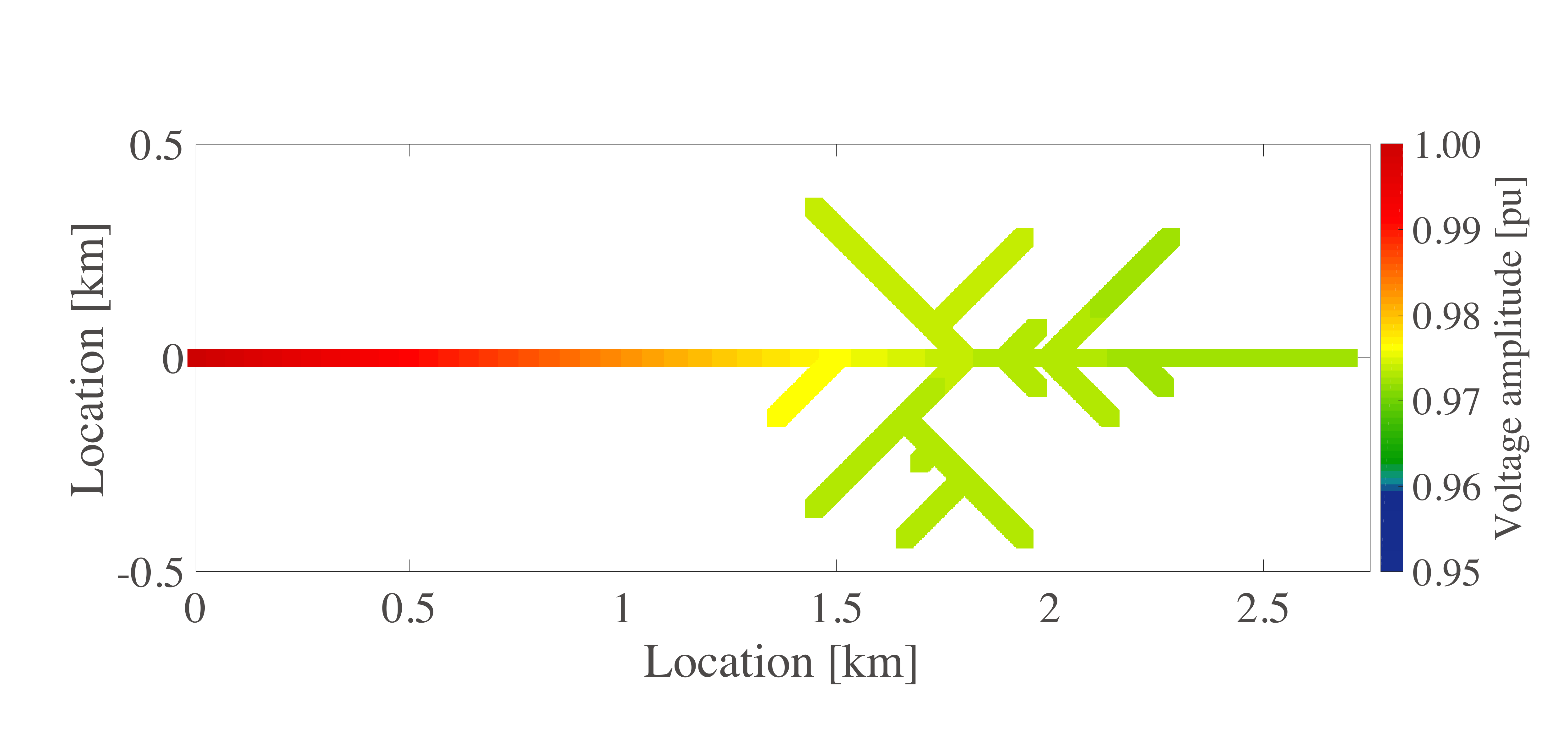}
\vspace{-4mm}
\subcaption{%
\footnotesize
Synthesized $P$ and $Q$
}%
\label{fig:rea_v_synth-simu}
\end{minipage}
\caption{%
Numerical simulations of the nonlinear ODE (\ref{eqn:ode}) incorporated with the power density functions based on Figures~\ref{fig:load2_19}, \ref{fig:syn_p_4}, and \ref{fig:rea_q_4}. 
The voltage amplitude at the starting of the feeder is set to unity that is a boundary condition of (\ref{eqn:ode}). 
The voltage gradient at each end of the feeders is set to zero that is another boundary condition. 
}%
\label{fig:rea_v_4}
\vspace{-3mm}
\end{figure*}

Figure~\ref{fig:rea_v_4} shows the numerical simulations of the nonlinear ODE (\ref{eqn:ode}) incorporated with the power density functions based on Figures~\ref{fig:syn_p_4} and \ref{fig:rea_q_4}. 
The simulations for the multiple feeders with bifurcations were performed with the same scheme as in \cite{Susuki2}. 
The voltage amplitude at the starting of the feeder is set to unity that is a boundary condition of (\ref{eqn:ode}). 
The voltage gradient at each open end of the feeders is set to zero that is another boundary condition. 
The simulations for the equally allocated pattern of active power is shown in Figure~\ref{fig:rea_v_equal-simu}. 
For this, we constructed the power density functions $p(x)$ and $q(x)$ with Figures~\ref{fig:load2_19}, \ref{fig:p_equal}, and \ref{fig:rea_q_equal}. 
The simulations for the synthesized patterns are shown in Figure~\ref{fig:rea_v_synth-simu} where we use $p(x)$ and $q(x)$ with Figures~\ref{fig:load2_19}, \ref{fig:p_synth}, and \ref{fig:rea_q_synth}. 
In comparison with Figure~\ref{fig:rea_v_equal-simu}, we vividly see in Figure~\ref{fig:rea_v_synth-simu} that the deviation of voltage gradient on the feeder surrounded by the \emph{dotted} line is greatly reduced, and thus that the voltage amplitude at each end of the feeders is increased. 
We conclude for the model of multiple feeders that the proposed algorithm does work for provision of AS (as in Tables~\ref{tab:pq_multiple_synth}) and mitigation of voltage impact. 

\begin{table}[t]
\centering
\scriptsize
\caption{%
Values of spatial patterns of the $16$ charging stations in the model of multiple feeders
}%
\subcaption{Uniform case in Figures~\ref{fig:p_equal} and \ref{fig:rea_q_equal}}
\label{tab:pq_multiple_equal}
\begin{tabular}{c||c|c|c|c|c|c} \hline
&St.\,1&St.\,2&St.\,3&St.\,4&St.\,5&St.\,6\\ \hline \hline
$P_i/10^{-4}$&$6.25$&$6.25$&$6.25$&$6.25$&$6.25$&$6.25$\\ \hline
$Q_i/10^{-4}$&$3.03$&$3.03$&$3.03$&$3.03$&$3.03$&$3.03$\\ \hline \hline
&St.\,7&St.\,8&St.\,9&St.\,10&St.\,11&St.\,12\\ \hline \hline
$P_i/10^{-4}$&$6.25$&$6.25$&$6.25$&$6.25$&$6.25$&$6.25$\\ \hline
$Q_i/10^{-4}$&$3.03$&$3.03$&$3.03$&$3.03$&$3.03$&$3.03$\\ \hline \hline
&St.\,13&St.\,14&St.\,15&St.\,16 & \multicolumn{2}{c}{Total} \\ \hline \hline
$P_i/10^{-4}$&$6.25$&$6.25$&$6.25$&$6.25$& \multicolumn{2}{c}{$\bf 100$}\\ \hline
$Q_i/10^{-4}$&$3.03$&$3.03$&$3.03$&$3.03$& \multicolumn{2}{c}{--}\\ \hline
\end{tabular}
\vspace*{5mm}
\subcaption{Synthesized case in Figures~\ref{fig:p_synth} and \ref{fig:rea_q_synth}}
\label{tab:pq_multiple_synth}
\begin{tabular}{c||c|c|c|c|c|c} \hline
&St.\,1&St.\,2&St.\,3&St.\,4&St.\,5&St.\,6\\ \hline \hline
$P_i/10^{-3}$&$-9.00$&$-9.00$&$-9.00$&$-9.00$&$-9.00$&$-6.33$\\ \hline
$Q_i/10^{-3}$&$3.01$&$-2.49$&$0.00$&$0.00$&$0.00$&$0.00$\\ \hline \hline
&St.\,7&St.\,8&St.\,9&St.\,10&St.\,11&St.\,12\\ \hline \hline
$P_i/10^{-3}$&$3.98$&$2.01$&$7.85$&$9.00$&$0.94$&$7.81$\\ \hline
$Q_i/10^{-3}$&$4.36$&$0.00$&$0.00$&$0.00$&$3.06$&$4.36$\\ \hline \hline
&St.\,13&St.\,14&St.\,15&St.\,16 & \multicolumn{2}{c}{Total}\\ \hline \hline
$P_i/10^{-3}$&$8.34$&$3.39$&$9.00$&$9.00$& \multicolumn{2}{c}{$\bf 10.0$}\\ \hline
$Q_i/10^{-3}$&$4.36$&$4.36$&$4.36$&$4.36$& \multicolumn{2}{c}{--}\\ \hline
\end{tabular}
\end{table}

\section{Conclusion}
\label{sec:conclusion}

We developed an algorithm for synthesizing the spatial pattern of charging/discharging operations of in-vehicle batteries for not only provision of AS but also mitigation of voltage impact. 
This algorithm is based on analysis and physical implications of the nonlinear ODE (\ref{eqn:ode}) for continuum representation of distribution voltage. 
It is easy for implementation in software and needs no iterative computation like optimization using power-flow equations, and thus is expected to be fast and scalable for large-scale applications. 
This is significant to the provision of primary frequency control reserve, which is the fastest responsiveness of AS. 
Effectiveness of the proposed algorithm was established with numerical simulations on the single feeder model and on the practical model of multiple feeders. 

Several remarks on the current work in this paper are presented. 
First, in this paper, we did not consider any voltage regulator (e.g. step voltage regulator) for the synthesis procedure. 
As shown in \cite{Susuki2}, since such a regulator is incorporated with the nonlinear ODE (\ref{eqn:ode}) by introducing a boundary condition, it is possible to use the regulator for the current synthesis purpose. 
Second, we assumed that the target feeder models have constant conductance and susceptance in position. 
This assumption can be relaxed substantially by introducing a boundary condition \cite{Susuki2}, while still applying the same analysis and algorithm as in this paper. 
Because the ratio of conductance and susceptance is particularly important with regard to the effectiveness of reactive-power compensation, it is necessary to appropriately determine the boundary condition at the point where the values of conductance and susceptance change. 
Third, in this paper, we considered the provision of AS by in-vehicle batteries under a static situation in time. 
But, when we provide AS as the primary frequency control reserve, 
it is necessary to evaluate the algorithm under dynamic situations of voltage and frequency.  
Regarding this, in this paper we derived the analytical solutions for constant power model.  
A remaining work is to derive analytical solutions for various load models and to establish a synthesis algorithm based on them.  
Finally, the achievement of AS provision by EVs highly depends on their primary use for transportation. 
Therefore, it is important to design the integrated transportation-energy management system that can handle actual data on vehicle location obtained with GPS or predicted data. 

\section*{Acknowledgment}

The second author greatly appreciates Mr. Seongchoel Baek and Professor Takashi Hikihara (Kyoto University) for their contributions to the initial research before the work presented here. 
The authors would like to thank Mr. Shota Yumiki (Osaka Prefecture University) for valuable discussions on the work and thank anonymous reviewers for their constructive comments.



\begin{thebibliography}{00}

\bibitem{Ipakchi} A. Ipakchi and F. Albuyeh, ``Grid of the future,'' IEEE Power \& Energy Mag., vol.\,7, issue\,2, pp.\,52--62, 2009.
\bibitem{Sousa} T. Sousa, H. Morais, Z. Vale, Senior, P. Faria, and J. Soares, ``Intelligent energy resource management considering vehicle-to-grid: A simulated annealing approach,'' IEEE Trans. on Smart Grid, vol.\,3, no.\,1, pp.\,535--542, 2012. 
\bibitem{Wu} C. Wu, H. Mohsenian-Rad, and J. Huang, ``Vehicle-to-aggregator interaction game,'' IEEE Trans. on Smart Grid, vol.\,3, no.\,1, pp.\,535--542, 2012. 
\bibitem{Arriaga} I. J. P. Arriaga, ``The transmission of the future,'' IEEE Power \& Energy Mag., vol.\,14, no.\,4, pp.\,41--53, 2016. 
\bibitem{Rebours} Y. G. Rebours, D. S. Kirschen, M. Trotignon, and S. Rossignol, ``A survey of frequency and voltage control ancillary services--Part I: Technical features,'' IEEE Trans. on Power Syst., vol.\,22, no.\,1, pp.\,350--357, 2007. 
\bibitem{Kempton} W. Kempton and J. Tomic, ``Vehicle-to-grid power fundamentals: Calculating capacity and net revenue,'' J. Power Sources, vol.\,144, no.\,1, pp.\,268--279, 2005. 
\bibitem{Tomic} J. Tomic and W. Kempton, ``Using fleets of electric-drive vehicles for grid support,'' J. Power Sources, vol.\,168, Issue.\,2, pp.\,459--468, 2007. 
\bibitem{Ota} Y. Ota, H. Taniguchi, T. Nakajima, K. M. Liyanage, J. Baba, and A. Yokoyama, ``Autonomous distributed {V2G} (Vehicle-to-Grid) satisfying scheduled charging,'' IEEE Trans. on Smart Grid, vol.\,3, no.\,1, pp.\,559--564, 2012.  
%
{\color{black}
\bibitem{Marinelli_JES7} M. Marinelli, S. Martinenas, K. Knezovi\'c, and P. B. Andersen, ``Validating a centralized approach to primary frequency control with series-produced electric vehicles," J. Energy Storage, vol.\,7, pp.\,63--73, 2016.}
%
\bibitem{Clement1} K. Clement-Nyns, E. Haesen, and J. Driesen, ``The impact of charging plug-in hybrid electric vehicles on residential distribution grid,'' IEEE Trans. on Power Syst., vol.\,25, no.\,1, pp.\,371--380, 2010. 
\bibitem{Clement2} K. Clement-Nyns, E. Haesen, and J. Driesen, ``The impact of vehicle-to-grid on the distribution grid,'' Electr. Power Syst. Res., vol.\,81, pp.\,185--192, 2011. 
\bibitem{Falahi} M. Falahi, H. M. Chou, M. Ehsani, L. Xie, and K. L. Butler-Purry, ``Potential power quality benefits of electric vehicles,'' IEEE Trans. on Sust. Energy, vol.\,4, no.\,4, pp.\,1016--1023, 2016. 
\bibitem{Itoh} A. Itoh, A. Kawashima, T. Suzuki, S. Inagaki, and T. Yamaguchi, ``Model predictive charging control of in-vehicle batteries for home energy management based on vehicle state prediction,'' IEEE Trans. on Control Sys. Tech., vol.\,26, issue\,1, pp.\,51--64, 2018.
\bibitem{Fairley} P. Fairley, ``Car sharing could be the EV's killer app?," IEEE Spectrum, vol.\,50, issue.\,9, pp.\,14--15, 2013. 
\bibitem{Tan} K. K. Tan, K. K. K. Htet, and A. S. Narayanan, ``Mitigation of vehicle distribution in an {EV} sharing scheme for last mile transportation,'' IEEE Trans. on Intelligent Transportation Syst., vol.\,16, no.\,5, pp.\,2631--2641, 2015. 
\bibitem{Kawashima:CCTA17} A. Kawashima, N. Makino, S. Inagaki, T. Suzuki, and O. Shimizu, ``Simultaneous optimization of assignment, reallocation and charging of electric vehicles in sharing services," in Proc. IEEE 1st Conference on Control Technology and Applications, pp.\,1070--1076, 2017. 
\bibitem{Autolib} Autolib' Paris, https://www.autolib.eu/en/. 
%
\bibitem{Chertkov} M. Chertkov, S. Backhaus, K. Turtisyn, V. Chernyak, and V. Lebedev, ``Voltage collapse and ODE approach to power flows: Analysis of a feeder line with static disorder in consumption/production,'' arXiv:1106.5003, 2011. 
\bibitem{Susuki2} Y. Susuki, N. Mizuta, A. Kawashima, Y. Ota, A. Ishigame, S. Inagaki, and T. Suzuki, ``A continuum approach to assessing the impact of spatio-temporal {EV} charging to power distribution grids,'' in Proc. IEEE 20th International Conference on Intelligent Transportation Systems, pp.\,2372--2377, 2017. 
\bibitem{Mizuta} N. Mizuta, Y. Susuki, Y. Ota, and A. Ishigame, ``An ODE-based design of spatial charging/discharging patterns of in-vehicle batteries for provision of ancillary service,'' in Proc. IEEE 1st Conference on Control Technology and Applications, pp.\,193--198, 2017. 
\bibitem{Machowski} J. Machowski, J.W. Bialek, and J.R. Bumby, ``Power System Dynamics and Stability,'' John Wiley \& Sons, 1997. 
\bibitem{Yang} Y. Yang, H. Wang, F. Blaabjerg, and T. Kerekes, ``A Hybrid Power Control Concept for PV Inverters With Reduced Thermal Loading,'' IEEE Trans. on Power Elec., vol.\,29, no.\,12, pp.\,6271--6275, 2014. 
\bibitem{Keane} A. Keane, L. F. Ochoa, E. Vittal, C. J. Dent, and G. P. Harrison, ``Enhanced utilization of voltage control resources with distributed generation,'' IEEE Trans. on Power Syst., vol.\,26, issue\,1, pp.\,252--260, 2011. 
\bibitem{Tanaka} H. Tanaka, T. Tanaka, T. Wakimoto, E. Hiraki, and M. Okamoto, ``Reduced-capacity smart charger for electric vehicles on single-phase three-wire distribution feeders with reactive power control,'' IEEE Trans. on Ind. Appl., vol.\,51, issue\,1, pp.\,315--324, 2015. 
\bibitem{Shimizu} K. Shimizu, T. Masuta, Y. Ota, and A. Yokoyama, ``Load frequency control in power system using vehicle-to-grid system considering the customer convenience of electric vehicles,'' in Proc. POWERCON, pp.\,1--8, 2010. 
\bibitem{Masuta} T. Masuta and A. Yokoyama, ``Supplementary load frequency control by use of a number of both electric vehicles and heat pump water heaters,'' IEEE Trans. on Smart Grid, vol.\,3, issue\,3, pp.\,1253--1262, 2012. 
\end{thebibliography}
\end{document}